\documentclass[a4paper,12pt]{article}

\usepackage{geometry}
\usepackage{amsthm}
\usepackage{amsmath}
\usepackage{amssymb}
\usepackage{enumerate}
\usepackage{mathrsfs}
\usepackage{multirow}
\usepackage{fullpage}
\usepackage{color}

\def\eqn{equation}
\def\cond{condition}
\def\tfn{transformation}
\def\comp{component}

\def\bckg{background}
\def\sm{sigma model}

\def\dd{Drinfeld double}

\def\4diml{four-dimensional}
\def\-1{^{-1}}
\def\half{\frac{1}{2}}
\def\coor{coordinate}

\def \unit {{\bf 1}}

\def\wh{\widehat}

\def\sm{sigma model}

\def\tbf{torsionless $B$-field}
\def\sgp{subgroup}

\def\cf{{\mathcal {F}}}
\def\ce{{\mathcal {E}}}
\def\cb{{\mathcal {B}}}

\def\ghat{\hat g}

\newcommand{\mez}{\hspace{15pt}}

\begin{document}


\title{Plane-parallel waves as duals of the flat background III: T-duality with torsionless $B$-field}

\author{Ladislav Hlavat\'y\footnote{ladislav.hlavaty@fjfi.cvut.cz}\\{\em Faculty of Nuclear Sciences and Physical Engineering},\\
Ivo Petr\footnote{ivo.petr@fit.cvut.cz}\\{\em Faculty of Information Technology},\\
Filip Petr\'asek\footnote{filip.petrasek@fjfi.cvut.cz}\\{\em Faculty of Nuclear Sciences and Physical Engineering}\\ Czech Technical University in Prague
}

\maketitle

\begin{abstract}
By addition of non-zero, but torsionless $B$-field, we expand the classification of (non-)Abelian T-duals of the flat background in four dimensions with respect to one$\text{-}$, two$\text{-}$, three$\text{-}$, and four$\text{-}$dimensional subgroups of Poincar\'e group. 
{We discuss the influence of the additional $B$-field on the
process of dualization and identify essential
parts of the \tbf{}  that cannot be eliminated {in general} by
\coor{} or gauge \tfn{} of the dual \bckg. These effects are
demonstrated using particular examples. Due to their physical
importance, we focus on duals whose metrics represent plane-parallel
waves. Besides the previously found metrics, we find new pp-waves
depending on parameters originating from the torsionless $B$-field.
These pp-waves are brought into their standard forms in Brinkmann
and Rosen coordinates.}


\end{abstract}

\tableofcontents


\section{Introduction}

In string theory, the dynamics of a string propagating in curved
background can be naturally characterized by non-linear sigma model
on a manifold $M$ given by a tensor field $\cf$ and action
\begin{equation}\label{sigmamodel}
S_{\cf}[\phi]=\int_\Omega
\partial_-\phi^{\mu}\cf_{\mu\nu}(\phi)\partial_+ \phi^\nu\,d\xi_+
d\xi_-.
\end{equation}
The symmetric and antisymmetric part of the tensor $\cf$ represent
metric ${\cal E}$ and torsion potential ${\cal B}${, called also
Kalb--Ramond field or simply $B$-field, of the \sm{} background.}

{One may apply Abelian \cite{buscher:ssbfe} or non-Abelian \cite{delaossa1992} T-duality \tfn{} to the sigma model 
 \eqref{sigmamodel} and obtain the dual sigma model whenever there is a Lie group $G$ generated by Killing vectors $K_a,\ a = 1, \ldots , \dim G$ whose action is free and leaves the tensor field $\cf$ invariant.} 
(Non-)Abelian T-duality turns out to be a special case of Poisson--Lie T-duality \cite{klise} and it is well-known that duals exist even for backgrounds without isometries. Nevertheless, despite the fact that the geometry behind T-duality is studied thoroughly (see e.g. \cite{CavGual,severa2015}  for the formulation in terms of Courant algebroids), the presence of symmetries remains crucial if one wants to dualize a particular background \cite{Bouwknegt2017}. 

{Renewed interest in (non-)Abelian T-duality was ignited by papers \cite{Hassan,1012.1320 sfethomp}, where it was extended to sigma models with non-vanishing Ramond fluxes, thus allowing to search for new supergravity solutions \cite{1104.5196 LCST,Itsios:2013wd}. {A special pp-wave/SYM duality is reviewed in \cite{SadSheikh}.  As the underlying structure accommodating T-duality is the \dd, it is also frequently used in the study of integrable sigma models and their deformations, see e.g. \cite{BorWulff1,BorWulff}} and references therein. However, our interest is slightly different. 

Finding solution of equations of motion
\begin{equation}\label{eqofmo}(\cf_{\mu\nu}+\cf_{\nu\mu})\,\partial_{+}\partial_{-}\phi^{\nu} +
  (\cf_{\mu\nu,\beta} + \cf_{\beta\mu,\nu} - \cf_{\beta\nu,\mu})\,
   \partial_{-}\phi^{\beta}\partial_{+}\phi^{\nu} = 0
\end{equation}
of a sigma model in non-trivial background is usually quite
complicated and every solvable case attracts considerable attention. On the other hand, equations of motion of a sigma model in flat torsionless background are easy to solve. As the dual model may have prominently different curvature properties, one may utilize T-duality when searching for non-trivial solvable sigma models. Solutions of such mutually dual models are then related by T-duality transformation.}

In previous papers \cite{php} and \cite{hp}, we focused on the sigma model in background given by flat four-dimensional metric, and classified its T-duals with respect to one-, two-, three-, and four-dimensional subgroups of Poincar\'e group {given in \cite{PWZ}. The duals were found using (non-)Abelian T-duality with and without spectators.} In both cases, we have considered the
flat background without $B$-field. However, besides such 
{model}, it makes sense to dualize 
{sigma models in backgrounds with flat metric and
non-zero, but torsionless $B$-field as the bulk equations of motion
\eqref{eqofmo} remain d'Alembert wave \eqn s.\footnote{Although, the
$B$-field may appear in boundary \cond s for open strings.}

In the following, we analyze the effects of additional $B$-field on
the process of dualization and on the properties of dual
backgrounds. In particular, we show that a suitable $B$-field allows
us to dualize with respect to subgroups that otherwise could not be
used for dualization, see \cite{php}. It is often possible to
partially eliminate the dependence of dual backgrounds on the
$B$-field using coordinate or gauge transformations. We
identify the relevant components of $\cb$ and point out the results
that could not be obtained with trivial $B$-field. {Through the whole paper, we address only local properties of sigma model backgrounds because global aspects of (non-)Abelian T-duality \cite{AAG:global} are still insufficiently understood and various issues remain obscure especially at the quantum level.}

As the title of paper suggests, we focus on dual backgrounds having pp-wave metric. These are often studied in context of string theory. Backgrounds of the pp-wave form provide exact solutions to string beta equations \cite{tsey:revExSol,tseyt94} and frequently emerge from WZW models \cite{tsey:revExSol,Sfetseyt94}. The ability to solve classical string equations allowed authors of papers \cite{BLPT,papa} to quantize the string propagating in certain classes of pp-wave backgrounds. Last but not least, pp-waves and their non-Abelian T-duals were investigated in context of supergravity e.g. in Refs. \cite{1205.2274 ILCS,DMRV,LNZ}.

The structure of paper is the following. In Sec. \ref{th_bckg}, we review necessary facts concerning the dualization procedure and pp-waves. Besides, we investigate thoroughly which parts of the torsionless $B$-field affect dual \bckg s. In Secs. \ref{1dimsubgps}, \ref{2dimsubgps}, \ref{3dimsubgps}, and \ref{4dimsubgps}, we find particular pp-waves as duals of the flat torsionless \bckg\ with emphasis on the cases where the addition of $B$-field provides new pp-waves. Dual pp-waves obtained before in Refs. \cite{php,hp} without $B$-field are summarized in the Appendix.

\section{Theoretical background}\label{th_bckg}

\subsection{Dualization with $B$-field}

{A thorough discussion concerning the process of dualization and the differences between atomic T-duality and T-duality with spectators has been given in \cite{php}. Therefore, we {do} not repeat the details, {but we} focus on the study of (non-)Abelian T-duality in the presence of non-zero, but torsionless $B$-field}.

{Consider sigma model \eqref{sigmamodel} on  four-dimensional target manifold $M$ for which there is a group of isometries represented by the Lie group $G$ with free action on $M$. The action of $G$ is transitive on its orbits, hence we may locally consider $M\approx (M/G) \times G$ and introduce adapted \coor s
\begin{equation}\label{adapted} x^\mu=\{s^\delta,g^a\},\ \
 \delta=1,\ldots,S,\ \ a=1,\ldots,{\dim}\,G, \ \ S=4-{\dim}\,G
\end{equation}
where $s^\delta$ label the orbits of $G$ and are treated as spectators during dualization. In these \coor s, dualizable tensor $\cf$ is defined by symmetric matrix $E_\xi(s)$ and antisymmetric matrix $B(s)$ as
\begin{equation}\label{met0}
        \left(\cf_\xi^B\right)_{\mu\nu}(s,g)=e_{\mu}^{j}(g)\left(E_\xi(s) +B(s)\right)_{jk}e_{\nu}^{k}(g), \qquad e_{\mu}^{j}(g)=\left(\begin{array}{cc} \unit_S & 0 \\ 0 &e_{c}^{a}(g) \end{array}\right)
\end{equation}
where $e_{c}^{a}(g)$ are components of the right-invariant Maurer--Cartan form $(dg)g^{-1}$.}

{We are interested in dualizable sigma models with flat background, where the metric (i.e. the symmetric part {${\cal E}$} of $\cf_\xi^B$) in suitable coordinates $x^{\rho}=(t, x, y, z)$ has the form $\eta_{\rho\sigma}=\text{diag}(-1,1,1,1)$, and its Killing vectors
\begin{equation}
    P_0=\partial_t,\quad P_j=\partial_j,\quad L_j=-\varepsilon_{jkl}x^k\partial_l,\quad
    M_j=-x^j\partial_t-t\partial_j
\end{equation}
generate the ten-dimensional Poincar\'e group containing $G$ as a subgroup. As derived in \cite{php}, it is not necessary to find the transformation between $(t, x, y, z)$ and the adapted coordinates to determine the {flat metric in the adapted \coor s \eqref{adapted}. Writing
\begin{equation}\label{Exiblok}
 E_\xi(s)=\left(\begin{array}{cc}
 E_{\alpha\beta}(s) & E_{\alpha d}(s) \\
 E_{c\beta}(s)      &E_{cd}(s)    \end{array}\right),\
 \ \alpha,\beta=1,\ldots,S,\ \ c,d=1,\ldots,\text{dim}\,G,
\end{equation}
 the components of the matrix $E_\xi(s)$ are} 
\begin{align}\nonumber
   E_{\alpha\beta}(s) &= \frac{\partial \xi^\rho}{\partial s^\alpha}(s)
   \, \eta_{\rho\sigma}\frac{\partial \xi^\sigma}{\partial s^\beta}(s),\\ \label{Exicomps}
  E_{\alpha d}(s) &=  \frac{\partial \xi^\rho}{\partial s^\alpha}(s)
   \, \eta_{\rho\sigma}K_d^\sigma(\xi(s)), \\\nonumber
  E_{c \beta}(s) &=
   K_c^\rho(\xi(s))\, \eta_{\rho\sigma}\frac{\partial \xi^\sigma}{\partial s^\beta}(s), \\\nonumber
 E_{cd}(s)&= K_c^\rho(\xi(s))
    \,\eta_{\rho\sigma}K_d^\sigma(\xi(s))
\end{align}
where {$K_a$ denote independent Killing vectors} generating the isometry group $G$, and the functions $\xi(s)$ assign the group unit of $G$ to points of orbits of $G$ in $M$. As suggested in \cite{hp:uniq}, $\xi(s)$ can be chosen arbitrarily up to certain conditions. For practical reasons, we choose these functions in accordance with \cite{php}.} 

{The matrix
\begin{equation}\label{Bcomponents}
B(s)=\left(\begin{array}{cc}
 B_{\alpha\beta}(s) & B_{\alpha d}(s) \\
 B_{c\beta}(s)      & B_{cd}(s)    \end{array}\right)=\left(
\begin{array}{cccc}
 0 & B_1(s) & B_2(s) & B_3(s) \\
 -B_1(s) & 0 & B_4(s) & B_5(s) \\
 -B_2(s) & -B_4(s) & 0 & B_6(s) \\
 -B_3(s) & -B_5(s) & -B_6(s) & 0 \\
\end{array}
\right)
\end{equation}
generates $G$-invariant $B$-field (i.e. the antisymmetric part $\cal B$ of $\cf_\xi^B$)} and torsion ${\cal H}=d\mathcal B$  via
\begin{equation}\label{Bfield}
{\cal B}=\half{\cal B}_{\mu\nu}(s,g)\,dx^\mu\wedge dx^\nu,\ \ {\cal B}_{\mu\nu}(s,g)=e_{\mu}^{j}(g)B(s)_{jk}e_{\nu}^{k}(g).
\end{equation}
Vanishing of the torsion restricts $B(s)$ in a way that depends on the choice of the Lie group $G$ with respect to which we dualize.

In case of (non-)Abelian T-duality, the dual tensor can be written as
\begin{equation}\label{Fghat2}
\wh \cf_\xi^B (s,\ghat)=(\unit+\wh E_\xi^B(s)\cdot\wh \Pi(\ghat))^{-1}\cdot \wh E_\xi^B(s)
\end{equation}
with matrix $\widehat E_\xi^B(s)$ given by
\begin{equation}\label{Extil}
  \wh E_\xi^B(s)=\big(D_1+\left(E_\xi(s)+B(s)\right)\cdot{D_2}
    \big)^{-1}\cdot\big(D_2+\left(E_\xi(s)+B(s)\right)\cdot D_1\big)
\end{equation}
where\begin{equation}\label{matsAB}
{D_1}=\left(\begin{array}{cc} \unit_S & 0 \\ 0 & \mathbf O_G
     \end{array}\right), \qquad
{D_2}=\left(\begin{array}{cc}  \mathbf O_S & 0 \\ 0 & \unit_G
     \end{array}\right) \end{equation}
and
\begin{equation} \label{Pihat2}
\wh \Pi(\hat g)=\left(\begin{array}{cc}  \mathbf O_S & 0 \\ 0 & -{f_{cd}}^{b}\ghat_b
  \end{array}\right)
\end{equation}
where ${f_{cd}}^{b}$ are structure coefficients of the Lie algebra
of the group  $G$ and $\ghat_b$ are coordinates of the Abelian group
$\wh G$.

As the first factor of $\wh E_\xi^B(s)$ in (\ref {Extil})  must have non-vanishing determinant, we get from \eqref{Exiblok}, \eqref{Bcomponents}, and \eqref{matsAB} condition of dualizability in the form
\begin{equation}\label{coofdu}
\det\, \left(D_1+\left(E_\xi(s)+B(s)\right)\cdot{D_2} \right)=\det\, \left(E_{cd}(s)+B_{cd}(s)\right)\neq 0.
\end{equation}
Compared with the case without $B$-field, the condition \eqref{coofdu} can be satisfied even for subgroups with vanishing $\det\, \left(E_{cd}(s)\right)$ labeled as not dualizable in \cite{php}. {Similar observation was made by Borsato and Wulff in the context of deformed T-dual models \cite{BorWulff}.} Consequently, the addition of advisable $B$-field may extend the number of
dualizable backgrounds.\footnote{Unfortunately, this does not work for one-dimensional subgroups where $B_{cd}(s)$ vanishes identically. However, it involves only one subgroup $S_{1,5}$.}

Using formulas \eqref{Fghat2}, \eqref{Extil}, and \eqref{Pihat2}, we gain dual tensors with symmetric and antisymmetric part that can be interpreted as dual metric and dual torsion potential.

\subsection{Reduction of $B$-field}\label{elimab}
{The dual backgrounds \eqref{Fghat2} may be very complicated in general. Nevertheless, the dependence of dual metric and torsion on the initial $B$-field can be often eliminated or simplified by coordinate or gauge \tfn.}

{The initial \bckg{} is described by 
the tensor \eqref{met0} that can be written as 
\begin{align}\nonumber
        \cf_\xi^B(s,g)=&\left(\begin{array}{cc}
 \cf_{\alpha\beta}(s,g) & \cf_{\alpha d}(s,g) \\
 \cf_{c\beta}(s,g)      &\cf_{cd}(s,g)    \end{array}\right)=\left(\begin{array}{cc}
 \ce_{\alpha\beta}(s,g)+\cb_{\alpha\beta}(s,g) & \ce_{\alpha d}(s,g)+\cb_{\alpha d}(s,g) \\
 \ce_{c\beta}(s,g)   +\cb_{c\beta}(s,g)      &\ce_{cd}(s,g)+\cb_{cd}(s,g)    \end{array}\right)
\\ =&\left(\begin{array}{cc}
 \delta_\alpha^\iota & 0 \\
0  &e_c^a(g) \end{array}\right)\cdot\left(\begin{array}{cc}
 E_{\iota\kappa}(s)+B_{\iota\kappa}(s) &  E_{\iota b}(s)+B_{\iota b}(s)  \\
  E_{a\kappa}(s)+B_{a\kappa}(s) & E_{ab}(s)+B_{ab}(s)  \end{array}\right)\cdot\left(\begin{array}{cc}
 \delta_\beta^\kappa & 0 \\
0  &e_d^b(g) \end{array}\right).
\label{matnab}
\end{align}
Vanishing of initial torsion $d\cb=0$ implies
\begin{align}\label{vanish torsion1}
    \partial_{[\alpha}\cb_{\beta\gamma]}&=0, & \partial_{[\alpha}\cb_{\beta]c}+\partial_{c}\cb_{\alpha\beta}&=0,\\
\label{vanish torsion2}
    \partial_{\alpha}\cb_{bc}+ \partial_{[b}\cb_{c]\alpha }&=0, & \partial_{[a}\cb_{b c]}&=0
\end{align}
where $X_{[AB]}:=\half(X_{AB}-X_{BA})$ and $X_{[ABC]}:=\frac{1}{3!}(X_{ABC}-X_{BAC}\pm\dots)$. Using \eqref{matnab}, we can rewrite \eqref{vanish torsion1} as
\begin{align}\label{vanish torsion3}
\partial_{[\alpha}B_{\beta\gamma]}&=0, & \partial_{[\alpha}B_{\beta]c}&=0.
\end{align}
The latter condition in \eqref{vanish torsion3} can be (at least locally) satisfied by
\begin{equation}\label{reseni B}B_{\alpha d}(s)=\partial_\alpha A_d(s).
\end{equation}
Using \eqref{matnab} and Maurer--Cartan equation 
\begin{equation}
\partial_c e_d^a-\partial_d e_c^a-{f_{jk}}^{a}e_c^j e_d^k=0, \label{cartan}
\end{equation} 
we can rewrite the former condition in \eqref{vanish torsion2} as
\begin{align}\label{vanish torsion4}
\partial_{\alpha}B_{cd}&={f_{cd}}^{b}B_{\alpha b}. 
\end{align}
Together with \eqref{reseni B} we get
\begin{equation}\label{reseni B2}
B_{cd}(s)=b_{cd}+{f_{cd}}^{b}A_b(s).
\end{equation}} 

{It is possible to derive general
coordinate \tfn s in the dual manifold that eliminate some components of the torsionless
$B$-field from the dual \bckg{}. Thus, we can neglect these \comp s and reduce the initial $B$-field.}

For better readability, we can introduce the block form of \eqref{matnab} in the following way
\begin{equation}\label{metabblock}
\cf_\xi^B(s,0)=\left(\begin{array}{cc}
 F_{SS} & F_{SG} \\
 F_{GS}      &F_{GG}    \end{array}\right)=\left(\begin{array}{cc}
 E_{SS}+B_{SS} & E_{SG}+B_{SG} \\
 (E_{SG})^T-(B_{SG})^T      &E_{GG}+B_{GG}     \end{array}\right).
\end{equation}
{Matrix of components of the dual tensor obtained from the equations \eqref{Fghat2}-\eqref{Pihat2} and \eqref{metabblock} can be written in the block form as  
\begin{equation}\label{dualFblock}
\wh\cf_\xi^B(s,\ghat)=\left(\begin{array}{cc}
\unit_S&-F_{SG} \\0&
\unit_G\end{array}\right)\cdot\left(\begin{array}{cc}
F_{SS}&0\\ 0&
\left[F_{GG}+{\wh \Pi_{GG}}\right]^{-1}\end{array}\right)\cdot\left(\begin{array}{cc}
\unit_{S}&0\\ 
F_{GS}&\unit_G\end{array}\right)
\end{equation} 
where $(\wh\Pi_{GG})_{cd}=-{f_{cd}}^{b}\ghat_b$. Using the \coor{} \tfn{} 
\begin{equation}\label{eliminateBdim2a}
 s^\alpha=s'^\alpha,\ \  \hat g_c=\hat g'_c+\Gamma_c(s'),
\end{equation}
we obtain the dual tensor \eqref{dualFblock} in the coordinates $(s',\ghat')$ as
$$ 
(\wh\cf_\xi^{B})'(s',\ghat')=\left(\begin{array}{cc}
\unit_S&\partial_{S}\Gamma_{G} \\0&
\unit_G\end{array}\right)\cdot\left(\begin{array}{cc}
 \wh \cf_{SS} & \wh \cf_{SG} \\
 \wh \cf_{GS}      &\wh \cf_{GG}    \end{array}\right)\cdot\left(\begin{array}{cc}
\unit_S&0 \\(\partial_{S}\Gamma_{G})^T& \unit_G\end{array}\right)=
$$
\begin{align}=\left(\begin{array}{cc}
\unit_S&\partial_{S}\Gamma_{G}-F_{SG} \\0&
\unit_G\end{array}\right)\cdot\left(\begin{array}{cc} F_{SS}&0\\ 0&
\left[F_{GG}+{\wh \Pi'_{GG}}\right]^{-1}\end{array}\right)\cdot\left(\begin{array}{cc}
\unit_{S}&0\\
(\partial_{S}\Gamma_{G})^T+F_{GS}&\unit_G\end{array}\right)
\label{tfn of F by Gamma},
\end{align}
where
$(\partial_{S}\Gamma_{G})_{\alpha d}=\partial_\alpha\Gamma_d(s')$ and $(\wh\Pi'_{GG})_{cd}=-{f_{cd}}^{b}\left(\ghat'_b+\Gamma_b(s')\right)$.
It is clear from \eqref{reseni B}, \eqref{reseni B2}, and \eqref{eliminateBdim2a} that by the choice $\Gamma_c(s')=A_c(s')$, \emph{we can
eliminate the dependence of the dual tensor \eqref{tfn of F by Gamma} on $A_c(s')$ by the coordinate transformation}. Besides, the dual torsion is not influenced by $B_{SS}$ due to the former condition in \eqref{vanish torsion3} and consequently, \emph{we can eliminate the dependence of the dual tensor \eqref{tfn of F by Gamma} on $B_{\alpha\beta}(s')$ by a gauge transformation}.}

{As result, we can assume without loss of generality that functions $A_c(s)$ and $B_{\alpha\beta}(s)$ vanish. Using \eqref{reseni B} and \eqref{reseni B2}, we then get $B_{\alpha d}(s)=0$ and $B_{cd}(s)=b_{cd}=const.$
It means that \emph{if we are interested in dual \bckg s only up to \coor{} and gauge \tfn s, we can restrict ourselves to initial $B$-fields given by the constant matrix }
\begin{equation} \label{B}
B(s)=\left(\begin{array}{cc}  \mathbf O_S & 0 \\ 0 & B_{cd}
  \end{array}\right),\quad B_{cd}=b_{cd}=const.
\end{equation}
The dual tensor \eqref{dualFblock} then reads
$$\wh\cf_\xi^{B}(s,\ghat)=\left(\begin{array}{cc}
\unit_S&-E_{SG} \\0&
\unit_G\end{array}\right)\cdot\left(\begin{array}{cc} E_{SS}&0\\ 0&
\left[E_{GG}+B_{GG}+{\wh \Pi_{GG}}\right]^{-1}\end{array}\right)\cdot\left(\begin{array}{cc}
\unit_{S}&0\\
(E_{SG})^T&\unit_G\end{array}\right),$$
where $(B_{GG})_{cd}=B_{cd}$. Owing to \eqref{cartan} and \eqref{B}, the latter condition in \eqref{vanish torsion2} can be rewritten as
\begin{equation}\label{vanish torsionA}
    \partial_{[a}\cb_{b c]}= (\partial_{[a}e_b^d)e_{c]}^hB_{dh}=-e_{[a}^je_b^ke_{c]}^h{f_{jk}}^d B_{dh}=0,
\end{equation}
resulting in the restricting condition for the constants $B_{cd}$ in the form
\begin{equation}\label{Brestriction}
{f_{[ab}}^d B_{c]d}=0.
\end{equation}

{Finally, let us emphasize that considerations made in this section are valid for arbitrary dimensions of $M$ and $G$ as well as for arbitrary initial metric.} { Constants $B_{cd}$ satisfying  \eqref{Brestriction}  can be considered as \comp s of a 2-cocycle of Lie algebra of the group $G$. Therefore, addition of the torsionless $B$-field to the initial background can always be interpreted as the application of the deformed T-duality \cite{BorWulff1}.\footnote{We are gratefull to the authors of \cite{BorWulff1} for pointing out this fact.}
}
\subsection{Plane-parallel waves}
In this paper, we restrict ourselves to duals of the flat torsionless background whose metrics are the so-called plane-parallel
(pp-)waves. 

In general, pp-wave metrics admit covariantly constant null Killing vector and exhibit particularly simple curvature properties since {their scalar curvature vanishes and in suitable coordinates the Ricci tensor} has only one non-zero component.

All pp-wave metrics obtained below can be expressed in the Brinkmann
coordinates as
\begin{equation}\label{BrinkMetrics}
   ds^{2}=\left(K_3(u)z_3^2+K_4(u)z_4^2\right)du^2 +2dudv +dz_3^2+dz_4^2,
\end{equation}
or the Rosen \coor s as
\begin{equation}\label{RosenMetrics}
   ds^{2}= 2dudv+C_1(u)dx_1^2+C_2(u)dx_2^2.
\end{equation}
In this case, there is rather simple relation between these two forms (see e.g. \cite{blauffpapa})
\begin{align}
C_1(u)=Q_1(u)^2,\qquad K_3(u)=\frac{1}{Q_1(u)}\frac{d^2Q_1}{du^2}(u),\label{BrinkmannRosen1}\\
C_2(u)=Q_2(u)^2,\qquad K_4(u)=\frac{1}{Q_2(u)}\frac{d^2Q_2}{du^2}(u). \label{BrinkmannRosen}
\end{align}
As one can see, the correspondence is not one-to-one because for
given $K_3(u)$ and $K_4(u)$ the functions $Q_1(u)$ and $Q_2(u)$ are
solutions of linear differential equations of the second order. In
other words, the Rosen form of metric is not unique in general.

{These pp-waves have the form of the Penrose--{G{\"u}ven} limit
\cite{Penrose,Gueven} with metric \eqref{BrinkMetrics} and torsion
\begin{equation}\label{Gueven_torsion}
    H=H(u)\,du\wedge dz_3\wedge dz_4.
\end{equation}
The one-loop conformal invariance conditions 
in this case simplify substantially to solvable differential equation
for the dilaton $\Phi=\Phi(u)$\begin{equation}\label{dilaton_eqn}
   \Phi''(u)+K_{3}(u)+K_{4}(u)+\frac{1}{2} H(u)^2=0.
\end{equation}}

Special pp-wave backgrounds obtained in Refs.
\cite{tsey:revExSol,Sfetseyt94} from gauged WZW models are given (up
to gauge \tfn{} of the $B$-field) in the Rosen coordinates as
\begin{equation}\label{sfets} ds^2= dudv + \frac{g_1 (u')}{ g_1 (u')g_2 (u)  + q^2 }\ dx_1^2 +\frac {g_2 (u)}{ g_1 (u')g_2(u)  + q^2} \ dx_2^2,
\end{equation}
$${\cal B}_{12}(u,v,x_1,x_2)= \frac{q }{  g_1 (u') g_2 (u) + q^2}$$
with the Sfetsos--Tseytlin parameters $q= const,\  u'=au + d \ $ ($a,d=const$) and $g_i{(u)}$ taking any of the following forms
\begin{equation}\label{g1g2}
\left\{ 1 ,\, u^2 ,\, \tanh^2 u ,\, \tan^2 u ,\,
 u^{-2} ,\, \coth^2 u ,\, \cot^2 u \right\}.
\end{equation}
Several of these pp-waves have been already discovered in Refs. \cite{php} and \cite{hp} {as duals of the flat background}. As we shall see, dualization with
non-zero, but torsionless $B$-fields yields also other cases
of \eqref{sfets}. 
{Exact string solutions from gauged WZW models in the form of plane gravitational waves and their duals are investigated also in the paper \cite{AntonObers}.}

In the following, we focus on duals of the flat torsionless backgrounds exhibiting pp-wave characteristics. Especially, we attend to subgroups that provide new pp-waves not found in \cite{php,hp} and find their Brinkmann and Rosen forms with the Sfetsos--Tseytlin parameters $g_1(u)\,,g_2(u),$ and $q$. We also cover some other attractive cases {that demonstrate general ideas developed in the previous section}. Discussion of pp-waves 
that appeared already in \cite{php,hp} without presence of the $B$-field is included in the Appendix.

\section{Duals with respect to one-dimensional subgroups}\label{1dimsubgps}

It is clear from above that torsionless initial \bckg s do not influence the T-duals with respect to the one-dimensional subgroups as $B_{cd}=0$.
Indeed, a torsionless background on $M$ is
obtained if the functions
 $B_i(s_1,s_2,s_3),\ i=1,\dots,6$ fulfill conditions (cf.\eqref{vanish torsion1} and \eqref{vanish torsion2})
\begin{equation}
\partial_{s_3} B_1(s_1,s_2,s_3)-\partial_{s_2}B_2(s_1,s_2,s_3)+\partial_{s_1}B_4(s_1,s_2,s_3)=0,\label{torsionless_Abel_1a}
\end{equation}
\begin{align}
\partial_{s_1}B_5(s_1,s_2,s_3)-\partial_{s_2}B_3(s_1,s_2,s_3)&=0,\nonumber\\
\partial_{s_1}B_6(s_1,s_2,s_3)-\partial_{s_3}B_3(s_1,s_2,s_3)&=0,\label{torsionless_Abel_1b}\\
\partial_{s_2}B_6(s_1,s_2,s_3)-\partial_{s_3}B_5(s_1,s_2,s_3)&=0\nonumber.
\end{align}
However, results of the Sec. \ref{elimab} imply that we can
eliminate $B_3, B_5, B_6$ by the coordinate \tfn{}
(\ref{eliminateBdim2a}) and $B_1, B_2, B_4$ by a gauge \tfn{}. It
means that \emph{addition of torsionless $B$-field does not {influence} the duals
with respect to one-dimensional \sgp s obtained in \cite{php}.}

\begin{table}
\begin{center}
\scriptsize {\renewcommand{\arraystretch}{1.4}
\begin{tabular}{|c || c | c |}
\hline
 & Generators & Free and transitive action for   \\
\hline \hline $S_{1,1}$ & $\begin{array}{c}
\cos \gamma L_3 + \sin\gamma M_3, \\
0 < \gamma < \pi, \gamma \neq \frac{\pi}{2}
\end{array}$ & $\begin{array}{c} t\neq 0 \lor x \neq 0 \ \lor \\ \lor\ y\neq 0 \lor z \neq 0 \end{array}$ \\
\hline
$S_{1,2}$ &  $L_3$ & $x\neq 0 \lor y \neq 0$  \\
\hline
$S_{1,3}$ & $M_3$ & $t\neq 0 \lor z \neq 0$  \\
\hline
$S_{1,4}$ & $L_2 + M_1$ & $x\neq 0 \lor t+z \neq 0$ \\
\hline
$S_{1,5}$ & $P_0 - P_3$ & $\mathbb{R}^4$ \\
\hline
$S_{1,6}$ & $P_3$ & $\mathbb{R}^4$ \\
\hline
$S_{1,7}$ & $P_0$ & $\mathbb{R}^4$ \\
\hline
$S_{1,8}$ & $\begin{array}{c} L_3 + \epsilon (P_0 + P_3),\\ \epsilon = \pm 1  \end{array}$ & $\mathbb{R}^4$ \\
\hline
$S_{1,9}$ & $\begin{array}{c} L_3 + \alpha P_0,\ \ \ \alpha > 0 \end{array}$ & $\mathbb{R}^4$ \\
\hline
$S_{1,10}$ & $\begin{array}{c} L_3 + \alpha P_3,\ \ \ \alpha \neq 0 \end{array} $ & $\mathbb{R}^4$ \\
\hline
$S_{1,11}$ & $\begin{array}{c} M_3 + \alpha P_1,\ \ \ \alpha > 0 \end{array}$ & $\mathbb{R}^4$ \\
\hline
$S_{1,12}$ & $L_2 + M_1 + P_0 + P_3$ & $\mathbb{R}^4$ \\
\hline
$S_{1,13}$ & $\begin{array}{c} L_2 + M_1 + \epsilon P_2, \\ \epsilon = \pm 1\end{array}$ & $\mathbb{R}^4$ \\
\hline
\end{tabular}
} \normalsize \normalsize \caption {{Generators of one-dimensional
subgroups of the Poincar\'e group with open subsets of transitive
and free action\label{table1}}}
\end{center}
\end{table}

Depending on the particular choice of {Killing vector generating group $G$}, we get various dual backgrounds. Duals obtained with respect to subgroups
generated by (see Tab. \ref{table1}) $S_{1,1}$, $S_{1,2}$,
$S_{1,3}$, $S_{1,8}$, $S_{1,9}$, $S_{1,10}$, $S_{1,11}$, $S_{1,12}$
do not represent pp-waves since their scalar curvatures do not
vanish. The initial background cannot be dualized with respect to
subgroup generated by $S_{1,5}$ because even after addition of
antisymmetric $B$-field the condition of dualizability
\eqref{coofdu} is not satisfied.

Duals obtained with respect to subgroups generated by $S_{1,6}$,
$S_{1,7}$ with $B$-field satisfying conditions
\eqref{torsionless_Abel_1a} and \eqref{torsionless_Abel_1b} become
flat backgrounds with vanishing torsion. The only pp-waves
obtained as duals of the flat metric and torsionless $B$-field might
appear via dualization with respect to subgroups generated by
$S_{1,4}$ and $S_{1,13}$. {Their form was already given in \cite{php}.}

We should emphasize that when the $B$-field is not torsionless, the
dual backgrounds may have completely different properties. As an
example, let us consider the case of $S_{1,4}$. If
\eqref{torsionless_Abel_1a} does not hold, the dual background is a
pp-wave with torsion. If \eqref{torsionless_Abel_1b} is violated,
the scalar curvature of the dual metric does not vanish, and it is
no longer a pp-wave.

\section{Duals with respect to two-dimensional subgroups}\label{2dimsubgps}

\begin{table}
\begin{center}
\scriptsize {\renewcommand{\arraystretch}{1.4}
\begin{tabular}{|c || c | c |}
\hline
 & Generators & Free and transitive action for \\
\hline \hline
$S_{2,1}$ &  $L_3, M_3$ & $\begin{array}{c} y z \neq 0 \lor x z \neq 0\ \lor \\
\lor\ t y \neq 0 \lor t x \neq 0\end{array}$ \\
\hline
$S_{2,2}$ & $L_2 + M_1, L_1-M_2$ & $t+z\neq 0$  \\
\hline
$S_{2,3}$ & $L_3, P_0-P_3$ & $x\neq 0 \lor y\neq 0$ \\
\hline
$S_{2,4}$ & $L_3, P_3$ & $x\neq 0 \lor y\neq 0$ \\
\hline
$S_{2,5}$ & $L_3, P_0$& $x\neq 0 \lor y\neq 0$ \\
\hline
$S_{2,6}$ & $M_3, P_1$ & $t\neq 0 \lor z\neq 0$ \\
\hline
$S_{2,7}$ & $L_2 + M_1, P_0 - P_3$ & $t+z\neq 0$ \\
\hline
$S_{2,8}$ & $L_2 + M_1, P_2$ & $t+z\neq 0 \lor x\neq 0$ \\
\hline
$S_{2,9}$ & $P_0-P_3, P_1$ & $\mathbb{R}^4$ \\
\hline
$S_{2,10}$ & $P_0, P_3$& $\mathbb{R}^4$ \\
\hline
$S_{2,11}$ & $P_1, P_2$ & $\mathbb{R}^4$ \\
\hline $S_{2,12}$ & $\begin{array}{c} L_2 + M_1, L_1-M_2+P_2
\end{array}$ & $\begin{array}{c}
y (t+z)\neq 0\  \lor \\ \lor\ x (1+t+z)\neq 0\ \lor \\
\lor\ (t+z) (1+t+z)\neq 0
\end{array}$ \\
\hline
$S_{2,13}$ & $\begin{array}{c} L_3+\epsilon(P_0+P_3), P_0-P_3,\\ \epsilon = \pm 1 \end{array}$ & $\mathbb{R}^4$ \\
\hline
$S_{2,14}$ & $\begin{array}{c} L_3+\alpha P_0, P_3,\ \ \ \alpha>0 \end{array}$ & $\mathbb{R}^4$ \\
\hline
$S_{2,15}$ & $\begin{array}{c}L_3+\alpha P_3, P_0,\ \ \ \alpha\neq 0 \end{array}$ & $\mathbb{R}^4$ \\
\hline
$S_{2,16}$ & $\begin{array}{c} M_3+\alpha P_2, P_1,\ \ \ \alpha>0 \end{array}$ & $\mathbb{R}^4$ \\
\hline
$S_{2,17}$ & $\begin{array}{c} L_2+M_1+P_0+P_3, P_0-P_3 \end{array}$ & $\mathbb{R}^4$ \\
\hline
$S_{2,18}$ & $\begin{array}{c} L_2+M_1+\epsilon P_2, P_0-P_3,\\ \epsilon = \pm 1 \end{array}$ & $\mathbb{R}^4$ \\
\hline
$S_{2,19}$ & $\begin{array}{c} L_2+M_1+P_0+P_3, P_2\end{array}$ & $\mathbb{R}^4$ \\
\hline
$S_{2,20}$ & $M_3, L_2+M_1$ & $\begin{array}{c} z (t+z)\neq 0\ \lor \\ \lor\ x (t+z)\neq 0\ \lor \\
\lor\ t (t+z)\neq 0\end{array}$ \\
\hline $S_{2,21}$ & $\begin{array}{c} \cos \gamma L_3 + \sin\gamma
M_3, P_0-P_3,\\ 0 < \gamma < \pi, \gamma \neq \frac{\pi}{2}
\end{array}$ & $\begin{array}{c} t+z\neq 0\ \lor \\ \lor\ x\neq 0
\lor y\neq 0 \end{array}$ \\
\hline
$S_{2,22}$ & $M_3, P_0-P_3$ & $t+z\neq 0$ \\
\hline $S_{2,23}$ & $\begin{array}{c} M_3+\alpha P_2, L_2+M_1,\ \ \ \alpha>0 \end{array}$ & $t+z\neq 0 \lor x\neq 0$ \\
\hline
$S_{2,24}$ & $\begin{array}{c} M_3+\alpha P_2, P_0-P_3,\ \ \ \alpha>0 \end{array}$ & $\mathbb{R}^4$ \\
\hline
\end{tabular}
\normalsize \caption { {Generators of two-dimensional subgroups of
the Poincar\'e group with open subsets of transitive and
free action}\label{table2}} }
\end{center}
\end{table}
Results of Sec. \ref{elimab} imply that we can eliminate $B_2,\dots,B_5$ by the  coordinate \tfn{} (\ref{eliminateBdim2a}) and $B_1$ by a gauge \tfn{}. Therefore, the only relevant component is the constant $B_6$ satisfying \eqref{Brestriction}.

Subgroups generated by $$S_{2,3},\ S_{2,7},\ S_{2,9},\ S_{2,18}$$ can be used for dualization only with  $B_6\neq 0$ because the condition \eqref{coofdu} for these subgroups reads $$\det\left(E_{cd}(s)+B_{cd}(s)\right)=B_6^2 \neq 0.$$ Setting $B_6=0$ in the other cases restores the results of \cite{php}.

Duals obtained with respect to subgroups generated by $S_{2,1}$, $S_{2,4}$, $S_{2,5}$, $S_{2,6}$, $S_{2,14}$, $S_{2,15}$, $S_{2,16}$, $S_{2,19}$, $S_{2,20}$, and $S_{2,23}$ do not represent pp-waves as their scalar curvatures do not vanish for any $B_6$. Duals obtained with respect to subgroups generated by $S_{2,7}$, $S_{2,9}$, $S_{2,10}$, $S_{2,11}$, $S_{2,17}$,  $S_{2,18}$, $S_{2,22}$, and $S_{2,24}$ and torsionless $B$-field
are flat backgrounds with vanishing torsion.

The other dual backgrounds are pp-waves. Nevertheless, the duals obtained via $S_{2,3}$, $S_{2,8}$, $S_{2,13}$, and $S_{2,21}$ reproduce pp-waves found previously without $B$-field. Thus, we defer the discussion of these backgrounds to the Appendix 
with the exception of the interesting case of $S_{2,3}$.

Below, we present duals with respect to subgroups generated by $S_{2,2}$ and $S_{2,12}$ where the introduction of $B$-field results in new pp-waves and the special case of $S_{2,3}$ that yields pp-wave only for a suitable $B$-field.

\subsection{Dualization with respect to subgroup generated by $S_{2,2}$}

Similarly to \cite{php}, we choose $\xi(s) = (-\frac{s_2}{2s_1},0,0,s_1+\frac{s_2}{2s_1})$ and get the dual background in the form $$\wh\cf(s,\hat g)=\left(
\begin{array}{cccc}
 1-\frac{s_2}{s_1^2} & \frac{1}{2
   s_1} & 0 & 0 \\
 \frac{1}{2 s_1} & 0 & 0 & 0 \\
 0 & 0 & \frac{s_1^2}{s_1^4+B_6^2}
   &- \frac{B_6}{s_1^4+B_6^2} \\
 0 & 0 & \frac{B_6}{s_1^4+B_6^2} &
   \frac{s_1^2}{s_1^4+B_6^2} \\
\end{array}
\right).$$
Subsequent coordinate transformation
\begin{align}s_1& =u, &  \hat g_1& =\frac{z_3
   \sqrt{B_6^2+u^4}}{u}, \nonumber \\  s_2& =-u^2+2 u v+\frac{ \left(B_6^2-u^4\right)
   \left(z_3^2+z_4^2\right)}{B_6^2+u^4}, & \hat g_2& =\frac{z_4 \sqrt{B_6^2+u^4}}{u} \nonumber
\end{align}
brings the dual metric to the Brinkmann form
$$ds^2=\frac{2u^2 \left(u^4-5 B_6^2\right)}{\left(u^4+B_6^2\right)^2}\left(z_3^2+z_4^2\right)du^2 + 2 du dv +dz_3^2 + dz_4^2.$$
The corresponding torsion of this model is $$
\wh {\cal H}=\frac{4 B_6 u}{u^4+B_6^2} du\wedge dz_3 \wedge dz_4.
$$
The Rosen form of the metric is
$$ ds^2= 2 du dv +\frac{u^2}{{u}^4+B_6^2}\,\left(dx_1^2 + dx_2^2\right)$$
that corresponds to the Sfetsos--Tseytlin parameters $g_1(u)=g_2(u)=u^2$ and $q=B_6$. This restores the result of \cite{php} for $B_6=0$. {For $B_6\neq 0$} \emph{we get new pp-wave}. {One can then} perform additional change of coordinates that effectively rescales $B_6$ to 1.

\subsection{Dualization with respect to subgroup generated by $S_{2,3}$}

 This is \emph{one of the cases where dualization is possible only for $B_6\neq 0$}. We choose $\xi(s) = (\frac{s_1}{2},s_2,0,\frac{s_1}{2})$ and get the dual background in the simple form $$\wh\cf(s,\hat g)=\left(
\begin{array}{cccc}
 -\frac{s_2^2}{B_6^2}-1 &0 &
   \frac{1}{B_6} &\frac{s_2^2}{B_6^2} \\
0& 1 & 0 & 0 \\
 \frac{1}{B_6} & 0 & 0 &-\frac{1}{B_6} \\
-\frac{s_2^2}{B_6^2} & 0 & \frac{1}{B_6} & \frac{s_2^2}{B_6^2} \\
\end{array}
\right). $$ 
The coordinate transformation
$$s_1=B_6\, u,\quad s_2=\sqrt{z_3^2+z_4^2},\quad \hat g_1=\frac{B_6^2 u}{2}+v,\quad \hat g_2=B_6 \arctan\frac{z_4}{z_3}$$
then brings the  dual  metric to Brinkmann form independent of the $B$-field
\begin{equation}\label{hom-isotr-pp}
ds^2=-\left(z_3^2+z_4^2\right)du^2 + 2 du dv +dz_3^2 + dz_4^2.
\end{equation}
The corresponding torsion is
\begin{equation}\label{hom-isotr-H}
\wh {\cal H}=2 du\wedge dz_3 \wedge dz_4.
\end{equation}
The Rosen form of the dual metric and torsion are
\begin{equation}\label{hom-Rosen}
ds^2 = 2 du dv +\cos^2 u\, dx_1^2 + \sin^2 u\, dx_2^2,
\end{equation}
\begin{equation}\label{hom-H Rosen}
\wh {\cal H}=\sin (2u) du\wedge dx_1 \wedge dx_2.
\end{equation}

This background appeared  already  in \cite{php,hp} by dualization with respect to subgroups generated by $S_{2,13},S_{2,21}, S_{4,17}$ and $S_{4,29}$ without $B$-field.

\subsection{Dualization with respect to subgroup generated by $S_{2,12}$}

We choose $\xi(s) = (-\frac{s_2}{1+s_1},0,0,s_1+\frac{s_2}{1+s_1})$ and get the dual background in the form
$$\wh\cf(s,\hat g)=\left(
\begin{array}{cccc}
 \frac{s_1^2+2 s_1-2 s_2+1}{(s_1+1)^2} & \frac{1}{s_1+1} & 0 & 0 \\
 \frac{1}{s_1+1} & 0 & 0 & 0 \\
 0 & 0 & \frac{(s_1+1)^2}{B_6^2+s_1^2 (s_1+1)^2} & -\frac{B_6}{B_6^2+s_1^2 (s_1+1)^2} \\
 0 & 0 & \frac{B_6}{B_6^2+s_1^2 (s_1+1)^2} & \frac{s_1^2}{B_6^2+s_1^2 (s_1+1)^2} \\
\end{array}
\right). $$
Subsequent coordinate transformation
$$s_1=u,$$
\begin{align*} s_2=&\frac{B_6^2 \left(-u^3+u^2 (2 v-1)+u \left(2 v+z_3^2+z_4^2\right)+z_4^2\right)}{2 u \left(B_6^2+u^2 (u+1)^2\right)}\\ & -\frac{u^2 (u+1)^2 \left(u^3+u^2 (1-2 v)+u \left(-2 v+z_3^2+z_4^2\right)+z_3^2\right)}{2 u \left(B_6^2+u^2 (u+1)^2\right)},
\end{align*}
$$\hat g_1=\frac{z_3 \sqrt{B_6^2+u^2 (u+1)^2}}{u+1},\qquad \hat g_2=-\frac{z_4 \sqrt{B_6^2+u^2 (u+1)^2}}{u}$$ brings this pp-wave metric to the Brinkmann form
$$ds^2=\left(\frac{2 u^2 (u+1)^4-\omega^2 (2 u (5 u+4)+1)}{\left(u^2 (u+1)^2+\omega^2\right)^2}z_3^2+\frac{2 u^4 (u+1)^2-\omega^2 (2 u (5 u+6)+3)}{\left(u^2 (u+1)^2+\omega^2\right)^2}z_4^2\right)du^2$$ $$ + 2 du dv +dz_3^2 + dz_4^2$$
where $\omega=B_6$.
The corresponding torsion of this model is $$
\wh {\cal H}=\frac{2 \omega (2 u+1)}{u^2 (u+1)^2+\omega^2} du\wedge dz_3 \wedge dz_4.
$$
The Rosen form of the metric is
$$ ds^2= 2 du dv +\frac{ \left(1+u\right)^2}{{u}^2
   (1+{u})^2+\omega^2}\,dx_1^2 +\frac{u^2}{{u}^2
   (1+{u})^2+\omega^2}\,dx_2^2$$
that corresponds to the Sfetsos--Tseytlin parameters $g_1(u')=(1+u)^2$, $g_2(u)=u^2$, and $q=\omega$. This restores the result of \cite{php} for $\omega=0$, but not so for general $\omega\neq0$. Moreover, the constant $\omega$ cannot be rescaled to a fixed value and hence, \emph{we get a new one-parametric set of pp-waves {dual to the flat torsionless background}.}

\section{Duals with respect to three-dimensional subgroups}\label{3dimsubgps}

\begin{table}
\begin{center}
\scriptsize{\renewcommand{\arraystretch}{1.4}
\begin{tabular}{|c||c|c|}
\hline
& Generators & Free and transitive action for \\
\hline \hline
$S_{3,1}$ &  $L_2+M_1, L_1-M_2, P_0-P_3$ & $t+z\neq0$ \\
\hline
$S_{3,2}$ & $L_3, P_0, P_3$ & $x\neq 0 \lor y\neq 0$ \\
\hline
$S_{3,3}$ & $M_3, P_1, P_2$ & $t\neq 0 \lor z\neq 0$ \\
\hline
$S_{3,4}$ & $L_2+M_1, P_0-P_3, P_2$ & $t+z\neq0$ \\
\hline
$S_{3,5}$ & $P_0-P_3, P_1, P_2$& $\mathbb{R}^4$ \\
\hline
$S_{3,6}$ & $P_1, P_2, P_3$ & $\mathbb{R}^4$ \\
\hline
$S_{3,7}$ & $P_0, P_1, P_2$ & $\mathbb{R}^4$ \\
\hline
$S_{3,8}$ & $L_2+M_1, L_1-M_2+P_2, P_0-P_3$ & $(t+z)(1+t+z)\neq0$ \\
\hline
$S_{3,9}$ & $L_2+M_1-\frac{1}{2}(P_0+P_3), P_0-P_3, P_2$ & $\mathbb{R}^4$ \\
\hline
$S_{3,10}$ & $M_3, L_2+M_1, P_2$ & $t+z\neq0$ \\
\hline
$S_{3,11}$ & $M_3, P_0-P_3, L_3$ & $(t+z)(x^2+y^2)\neq0$ \\
\hline
$S_{3,12}$ & $M_3, P_0-P_3, P_2$ & $t+z\neq0$ \\
\hline
{$S_{3,13}$} & $M_3 + \alpha P_2, P_0-P_3, P_1,\ \ \ \alpha > 0$ &{$\mathbb{R}^4$} \\
\hline
$S_{3,14}$ & $L_2+M_1, P_1, P_0-P_3$ & $\emptyset$ \\
\hline
{$S_{3,15}$} & $L_2+M_1, P_2+\beta P_1, P_0-P_3,\ \ \ \beta\neq 0$ & {$t+z\neq0$} \\
\hline
\multirow{2}{*}{$S_{3,16}$} & $L_2+M_1-\epsilon P_2, L_1-M_2+\beta P_2-\epsilon P_1, P_0-P_3,$ & \multirow{2}{*}{$\epsilon^2 + (t+z)(\beta+t+z)\neq0$} \\
 & $\beta > 0,\ \epsilon = \pm 1$ & \\
\hline
$S_{3,17}$ & $L_2+M_1-\epsilon P_2, L_1-M_2-\epsilon P_1, P_0-P_3, \ \ \ \epsilon = \pm 1$ & $\mathbb{R}^4$ \\
\hline
$S_{3,18}$ & $L_2+M_1-\frac{1}{2}(P_0+P_3), P_1, P_0-P_3$ & $\mathbb{R}^4$ \\
\hline
{$S_{3,19}$} & $L_2+M_1-\epsilon P_2,P_1,P_0-P_3,\ \ \ \epsilon = \pm 1$ & {$\mathbb{R}^4$} \\
\hline
$S_{3,20}$ & $L_2+M_1-\frac{1}{2}(P_0+P_3), P_2-\beta P_1, P_0-P_3, \ \ \ \beta \neq 0$ & $\mathbb{R}^4$ \\
\hline
\multirow{2}{*}{$S_{3,21}$} & $L_2+M_1-\epsilon P_2, P_2 - \beta P_1, P_0-P_3,$ & \multirow{2}{*}{$t+z+\beta\epsilon\neq0$} \\
& $\beta \neq 0,\ \epsilon = \pm 1$ & \\
\hline
{$S_{3,22}$} & $M_3 + \alpha P_1, L_2+M_1,P_0-P_3,\ \alpha > 0$ & {$t+z\neq0$} \\
\hline
\multirow{2}{*}{$S_{3,23}$} & $M_3 - \alpha P_2+\beta P_1, L_2+M_1, P_0-P_3,$ & \multirow{2}{*}{$t+z\neq0$} \\
 & $\alpha > 0,\ \beta \neq 0$ & \\
\hline
$S_{3,24}$ & $M_3, L_2+M_1,L_1-M_2$ & $t+z\neq0$ \\
\hline
$S_{3,25}$ & $M_3, L_2+M_1, P_0-P_3$ & $t+z\neq0$ \\
\hline
{$S_{3,26}$} & $M_3+\alpha P_2, L_2+M_1, P_0-P_3,\ \ \ \alpha>0$ & {$t+z\neq0$} \\
\hline
$S_{3,27}$ & $\cos \gamma L_3 + \sin\gamma M_3, P_0, P_3, \ \ \ 0 < \gamma < \pi, \gamma \neq \frac{\pi}{2}$ & $x\neq 0 \lor y\neq 0$ \\
\hline
$S_{3,28}$ & $M_3, P_0, P_3$ & $\emptyset$ \\
\hline
{$S_{3,29}$} & $M_3+\alpha P_2, P_0, P_3,\ \ \ \alpha>0$ & {$\mathbb{R}^4$} \\
\hline
$S_{3,30}$ &  $L_3, L_2+M_1, L_1-M_2$ & $\emptyset$ \\
\hline
{$S_{3,31}$} & $\cos \gamma L_3 + \sin\gamma M_3, P_1, P_2,\ \ \ 0 < \gamma < \pi, \gamma \neq \frac{\pi}{2}$ & {$t\neq 0 \lor z\neq 0$} \\
\hline
$S_{3,32}$ &  $L_3, P_1, P_2$ & $\emptyset$ \\
\hline
{$S_{3,33}$} &  $L_3+\epsilon(P_0-P_3), L_2+M_1, L_1-M_2,\ \epsilon = \pm 1$ & {$t+z\neq0$} \\
\hline
{$S_{3,34}$} &  $L_3-\epsilon(P_0+P_3), P_1, P_2,\ \epsilon = \pm 1$ & {$\mathbb{R}^4$} \\
\hline
{$S_{3,35}$} &  $L_3+\alpha P_0, P_1, P_2,\ \alpha >0$ & {$\mathbb{R}^4$} \\
\hline
{$S_{3,36}$} &  $L_3+\alpha P_3, P_1, P_2,\ \alpha \neq 0$ & {$\mathbb{R}^4$} \\
\hline
\multirow{2}{*}{$S_{3,37}$} & $\cos \gamma L_3 + \sin\gamma M_3, L_2+M_1, L_1-M_2,$ & \multirow{2}{*}{$t+z\neq0$} \\
 & $0 < \gamma < \pi, \gamma \neq \frac{\pi}{2}$ &   \\
\hline
$S_{3,38}$ & $L_3, M_1, M_2$ & $\emptyset$ \\
 \hline
$S_{3,39}$ & $L_1, L_2, L_3$ & $\emptyset$  \\
\hline
\end{tabular}
\normalsize \caption{Generators of three-dimensional subgroups of
the Poincar\'e group with open subsets of transitive and free action} \label{table3}}
\end{center}
\end{table}

{Results of Sec. \ref{elimab} imply that we can eliminate $B_1$,
$B_2$, and $B_3$ by the  coordinate \tfn{} \eqref{eliminateBdim2a}.
{From Eq. \eqref{B} we find} that the only relevant components
are the constants $B_4$, $B_5$, and $B_6$ satisfying
\eqref{Brestriction}.}

There are many subgroups, namely those generated by
$$S_{3,1},\ S_{3,4},\ S_{3,5},\ S_{3,8},\ S_{3,15},\ S_{3,16},\ S_{3,17},\ S_{3,19},\ S_{3,21},\ S_{3,33},\ S_{3,34}$$ (see Tab. \ref{table3}), that cannot be used for dualization without a $B$-field since $\det (E_{cd}(s))=0$ (see Tabs. 9-11 in \cite{php}). Addition of the suitable $B$-field with vanishing torsion enables dualization. 

Duals obtained with respect to subgroups generated by $S_{3,5}$, $S_{3,6}$, $S_{3,7}$, $S_{3,12}$, and $S_{3,13}$ with initial torsionless $B$-field are flat torsionless backgrounds.
Duals obtained with respect to subgroups generated by $S_{3,3}$, $S_{3,9}$, $S_{3,10}$, $S_{3,11}$, $S_{3,18}$, $S_{3,20}$, $S_{3,24}$, $S_{3,29}$, $S_{3,31}$, and $S_{3,33},\dots,S_{3,37}$ do not represent pp-waves as their scalar curvatures do not vanish for any initial torsionless $B$-field.
Since the actions of subgroups generated by $S_{3,14}$, $S_{3,28}$, $S_{3,30}$, $S_{3,32}$, $S_{3,38}$, and $S_{3,39}$ on $M$ are not transitive and free, they cannot be used for dualization.

{The other subgroups provide  pp-waves for suitable $B$-fields. Nevertheless, the duals obtained via $S_{3,1}$, $S_{3,2}$, $S_{3,4}$, $S_{3,8}$, $S_{3,15}$, $S_{3,16}$, $S_{3,17}$, $S_{3,19}$, $S_{3,21}$, $S_{3,22}$, $S_{3,23}$, $S_{3,25}$, $S_{3,26}$, and $S_{3,27}$ reproduce pp-waves found  previously without $B$-field. Thus, we defer the discussion of these backgrounds to the Appendix with the exception {of the case} of $S_{3,2}$ presented below.}

\subsection{Dualization with respect to subgroup generated by $S_{3,2}$}

This is rather interesting case since \emph{dualization with respect to the subgroup generated by $S_{3,2}$ produces
pp-wave only for} $B_6=\pm 1$, $B_4, B_5=const.,\ B_4^2 - B_5^2 \neq0$. We choose $\xi(s)=(0,\sqrt{s_1},0,0)$ with  $s_1 >0$.
Scalar curvature of the general dual background is
$$\wh {\cal R}=\frac{2 \left(B_6^2-1\right) \left(-5 B_4^2+5 B_5^2+2 \left(B_6^2-1\right)s_1\right)}{\left(B_4^2-B_5^2+\left(B_6^2-1\right) s_1\right)^2}.$$
Setting $B_6=\pm 1$ gives the dual background in the form
$$\wh{\cal F}(s,\hat g)=\left(
\begin{array}{cccc}
 \frac{1}{4 s_1} & 0 & 0 & 0 \\
 0 & 0 & -\frac{B_4+B_5 B_6}{B_4^2-B_5^2} & \frac{B_5+B_4 B_6}{B_4^2-B_5^2} \\
0 & \frac{B_4-B_5 B_6}{B_4^2-B_5^2} & \frac{B_5^2+s_1}{B_4^2-B_5^2} & -\frac{B_4 B_5+B_6
   s_1}{B_4^2-B_5^2} \\
 0 & \frac{B_5-B_4 B_6}{B_5^2-B_4^2} & \frac{B_6 s_1-B_4 B_5}{B_4^2-B_5^2} &
   \frac{B_4^2-s_1}{B_4^2-B_5^2} \\
\end{array}
\right).$$
Transformation of coordinates
\begin{align*}
s_1&=z_3^2+z_4^2, & \hat{g}_2&=-B_5B_6 u + B_4 \arctan\frac{z_4}{z_3},\\
\hat{g}_1&=-\frac{1}{2} \left(2v+B_5^2 u-B_4^2 u\right), & \hat{g}_3&=-B_4B_6 u + B_5 \arctan\frac{z_4}{z_3}
\end{align*}
eliminates the dependence on the $B$-field and we get the metric \eqref{hom-isotr-pp} and torsion \eqref{hom-isotr-H}. 

\section{Duals with respect to four-dimensional subgroups}\label{4dimsubgps}

\begin{table}
\begin{center}
\scriptsize{\renewcommand{\arraystretch}{1.4}
\begin{tabular}{|c||c|c|c|}
\hline
& Generators & Free and transitive action for \\
\hline \hline
$S_{4,1}$ & $P_0, P_1, P_2, P_3$  & $\mathbb{R}^4$ \\
\hline
$S_{4,2}$ & $M_3, P_0 - P_3, P_1, P_2$ & $t+z\neq0$ \\
\hline
$S_{4,3}$ & $L_1-M_2, P_2, P_0-P_3,L_2+M_1$ & $\emptyset$ \\
\hline
$S_{4,4}$ & $L_2+M_1, P_1,P_0-P_3,P_2$ & $\emptyset$ \\
\hline
$S_{4,5}$ & $\begin{array}{c} L_1-M_2-\epsilon P_1, P_2, P_0-P_3, L_2+M_1+\epsilon P_2,\\ \epsilon =\pm 1 \end{array}$  & $\emptyset$ \\
\hline
$S_{4,6}$ & $L_2 + M_1-\half (P_0+P_3),P_1, P_0 - P_3,P_2$ & $\mathbb{R}^4$ \\
\hline
$S_{4,7}$ & $\begin{array}{c} 2  M_3 + \alpha\,P_1, L_2 + M_1,  P_0 - P_3, P_2,\ \  \ \alpha >0 \end{array}$ & $t+z\neq0$ \\
\hline
$S_{4,8}$ & $M_3 , L_2 + M_1,  P_0 - P_3, P_2$ & $t+z\neq0$ \\
\hline
$S_{4,9}$ & $M_3, P_0, P_3, L_3$ & $\emptyset$ \\
\hline
$S_{4,10}$ & $M_3, P_0, P_3, P_1$ & $\emptyset$ \\
\hline
$S_{4,11}$ & $\begin{array}{c} M_3 + \alpha\,P_2,P_0,  P_3, P_1,\ \ \  \alpha >0 \end{array}$ & $\mathbb{R}^4$ \\
\hline
$S_{4,12}$ & $L_3, L_1-M_2, L_2+M_1, P_0-P_3$ & $\emptyset$ \\
\hline
$S_{4,13}$ & $L_3, P_1, P_2, M_3$  & $\emptyset$ \\
\hline
$S_{4,14}$ & $L_3, P_1, P_2, P_0-P_3$ & $\emptyset$ \\
\hline
$S_{4,15}$ & $L_3, P_1, P_2, P_3$ & $\emptyset$ \\
\hline
$S_{4,16}$ & $L_3, P_1, P_2, P_0$ & $\emptyset$ \\
\hline
$S_{4,17}$ & $\begin{array}{c} L_3+ \epsilon\,(P_0+ P_3), P_1, P_2, P_0- P_3,\ \ \ \epsilon =\pm 1 \end{array}$ & $\mathbb{R}^4$ \\
\hline
$S_{4,18}$ & $\begin{array}{c} L_3 + \alpha\,P_0, P_1, P_2, P_3,\ \ \ \alpha >0 \end{array}$ & $\mathbb{R}^4$ \\
\hline
$S_{4,19}$ & $\begin{array}{c} L_3 + \alpha\,P_3, P_1, P_2, P_0,\ \ \ \alpha \neq 0 \end{array}$ & $\mathbb{R}^4$ \\
\hline
$S_{4,20}$ & $M_1, M_2, L_3, P_3$ & $\emptyset$ \\
\hline
$S_{4,21}$ & $L_1, L_2, L_3, P_0$ & $\emptyset$ \\
\hline
$S_{4,22}$ & $L_2+M_1, P_0+P_3, P_0-P_3, P_1$  & $\emptyset$ \\
\hline
$S_{4,23}$ & $\begin{array}{c} L_2 + M_1-\half (P_0+P_3),L_1-M_2+\alpha\,P_1, P_0 - P_3, P_2,\\ \alpha \neq 0 \end{array}$ & $\mathbb{R}^4$ \\
\hline
$S_{4,24}$ & $L_2+M_1-\half(P_0+P_3), L_1-M_2, P_0-P_3, P_2$ & $\emptyset$ \\
\hline
$S_{4,25}$ & $\begin{array}{c} L_2 + M_1-\epsilon P_2, P_0 + P_3,P_1, P_0 - P_3,\\ \epsilon =\pm 1 \end{array}$ & $\mathbb{R}^4$ \\
\hline
$S_{4,26}$ & $\begin{array}{c} M_3 + \alpha\,P_1, L_2 + M_1, L_1 - M_2, P_0 - P_3,\\ \alpha >0 \end{array}$ & $t+z\neq0$ \\
\hline
$S_{4,27}$ & $M_3, L_1 - M_2, L_2 + M_1, P_0 - P_3$ & $t+z\neq0$ \\
\hline
$S_{4,28}$ & $\begin{array}{c} L_3 - \tan \gamma \,M_3, L_1 - M_2, L_2 + M_1, P_0 - P_3,\\ 0< \gamma < \pi, \gamma \neq \frac{\pi}{2} \end{array}$ & $t+z\neq0$ \\
\hline
$S_{4,29}$ & $\begin{array}{c} L_3 - \tan \gamma \,M_3, P_0 - P_3, P_1, P_2,\\ 0< \gamma < \pi, \gamma \neq \frac{\pi}{2} \end{array}$ & $t+z\neq0$ \\
\hline
$S_{4,30}$ & $M_3, P_1, P_0-P_3, L_2+M_1$ & $\emptyset$ \\
\hline
$S_{4,31}$ & $\begin{array}{c} M_3, P_1 + \beta\,P_2, P_0 - P_3, L_2 + M_1,\\ \beta \neq 0 \end{array}$ & $t+z\neq0$ \\
\hline
$S_{4,32}$ & $\begin{array}{c} M_3 + \alpha\,P_2, P_1, P_0 - P_3, L_2 + M_1,\\ \alpha > 0, \beta \neq 0 \end{array}$  & $\emptyset$ \\
\hline
$S_{4,33}$ & $\begin{array}{c} M_3 + \alpha\,P_2, P_1 + \beta\,P_2, P_0 - P_3, L_2 + M_1,\\ \alpha > 0, \beta \neq 0 \end{array}$  & $t+z\neq0$ \\
\hline
$S_{4,34}$ & $\begin{array}{c} L_3, L_2+M_1-\epsilon P_2, L_1-M_2-\epsilon P_1, P_0-P_3,\\ \epsilon =\pm 1 \end{array}$ & $\emptyset$ \\
\hline
$S_{4,35}$ & $L_3, K_3, L_2+K_1, L_1-K_2$  & $\emptyset$ \\
\hline
\end{tabular}
\normalsize \caption{Generators of four-dimensional subgroups of the Poincar\'e group with open subsets of transitive and free action} \label{table4}}
\end{center}
\end{table}

There are no spectators in this case and the $B$-field is calculated using a constant matrix $B$, see \eqref{Bfield}. Since the actions of subgroups generated by $S_{4,3}$, $S_{4,4}$, $S_{4,5}$, $S_{4,9}$, $S_{4,10}$, $S_{4,12}, \dots, S_{4,16}$, $S_{4,20}, \dots, S_{4,22}$, $S_{4,24}$, $S_{4,30}$,  $S_{4,32}$, $S_{4,34}$ and $S_{4,35}$ on $M$ are not transitive and free, these subgroups cannot be used for dualization (see Tab. \ref{table4}). On the other hand, for all the other subgroups the determinant $\det (E_{cd}(s))$ does not vanish, and dualization is possible even with a trivial $B$-field.

Duals obtained with respect to subgroups generated by $S_{4,1}$ and $S_{4,2}$ with initial torsionless $B$-field are flat torsionless backgrounds. Dualization with respect to $S_{4,6}$ and $S_{4,11}$ gives us backgrounds where a suitable choice of constants $B_i$ leads to vanishing scalar curvature. However, the resulting backgrounds then become flat and torsionless.

Duals obtained with respect to subgroups generated by $S_{4,18}$ and $S_{4,19}$ do not represent pp-waves as their scalar curvatures do not vanish for any initial torsionless $B$-field.

The other subgroups provide  pp-waves for suitable $B$-fields. Nevertheless, the duals obtained via $S_{4,7}$, $S_{4,8}$, $S_{4,17}$, $S_{4,23}$, $S_{4,25}$, $S_{4,26}$, $S_{4,27}$, $S_{4,28}$, $S_{4,29}$, $S_{4,31}$ and $S_{4,33}$ reproduce pp-waves found previously in the paper \cite{hp} without $B$-field. In other words, \emph{no new pp-waves are found by addition of the torsionless $B$-field to the flat metric}. We defer the discussion of these pp-wave {backgrounds to 
the Appendix}.

\section{Conclusion}

We have investigated pp-wave backgrounds that can be found as
(non-)Abelian T-duals of the flat metric supplemented by torsionless
$B$-fields.  It completes, together with papers \cite{php,hp}, the
classification of pp-waves obtainable as (non-)Abelian T-duals of
the  flat Lorentzian torsionless background in four dimensions with
respect to one$\text{-}$, two$\text{-}$, three$\text{-}$, and
four$\text{-}$dimensional subgroups of the Poincar\'e group listed
in \cite{PWZ}. Results are summarized in Tab. \ref{tableconcl1}.

There are several subgroups, namely
$$S_{2,3},\ S_{2,7},\ S_{2,9},\ S_{2,18},$$
$$S_{3,1},\ S_{3,4},\ S_{3,5},\ S_{3,8},\ S_{3,15},\ S_{3,16},\ S_{3,17},\ S_{3,19},\ S_{3,21},\ S_{3,33},\ S_{3,34}$$
that can be used for dualization only in {presence} of advisable non-trivial $B$-field since the condition of dualizability \eqref{coofdu} must be satisfied. All the corresponding duals might be interpreted as complement of the classification performed in \cite{php}. Nevertheless, we have investigated only those that produce pp-wave backgrounds.

{In general considerations made in Sec. \ref{elimab}, we
have identified parts of the \tbf{} that influence results of the dualization,
i.e., those that cannot be eliminated {in principle}  by \coor{} or gauge \tfn{} of
the dual \bckg. This 
significantly simplified computations made in the following sections as we could omit several \comp s of the
initial $B$-field or set them to constants.}

Plane-parallel waves can be gained by dualization with initial torsionless $B$-field with respect to subgroups generated by (see Tabs. \ref{table1}-\ref{table4})
$$
S_{1,4},\ S_{1,13},
$$
$$
S_{2,2},\ S_{2,3},\ S_{2,8},\ S_{2,12},\ S_{2,13},\ S_{2,21},
$$
$$
S_{3,1},\ S_{3,2},\ S_{3,4},\ S_{3,8},\ S_{3,15},\ S_{3,16},\ S_{3,17},\ S_{3,19},\ S_{3,21},\ S_{3,22},\ S_{3,23},\ S_{3,25},\ S_{3,26},\ S_{3,27},
$$
$${
S_{4,7},\ S_{4,8},\ S_{4,17},\ S_{4,23},\ S_{4,25},\ S_{4,26},\ S_{4,27},\ S_{4,28},\ S_{4,29},\ S_{4,31},\ S_{4,33}.}
$$
All found pp-waves are of the Sfetsos--Tseytlin type, i.e., they can be expressed in the form \eqref{sfets} with appropriate parameters $g_1(u')$, $g_2(u)$, and $q$.

{It turns out} that in several cases, namely for the subgroups generated by
\begin{equation}\label{restore}
\begin{array}{c}
S_{1,4},\ S_{1,13},\ S_{2,13},\ S_{2,21},\ S_{3,22},\ S_{3,23},\ S_{3,25},\ S_{3,26},\\
{S_{4,7},\ S_{4,8},\ S_{4,17},\ S_{4,23},\ S_{4,25},\ S_{4,26},\ S_{4,27},\ S_{4,28},\ S_{4,29},\ S_{4,31},\ S_{4,33},}
\end{array}
\end{equation}
the dependence of dual backgrounds on $B$-fields can be completely eliminated by coordinate and gauge  transformations. It means that the addition of the $B$-field does not affect the dual model and it restores the result of {\cite{php,hp}} found previously.

{Dualization with respect to} subgroups generated by
\begin{equation}\label{oneknown}
S_{2,3},\ S_{3,1},\ S_{3,17},\ S_{3,21},
\end{equation}
\begin{equation}\label{moreknown}
{S_{2,8},\ S_{3,4},\ S_{3,8},\ S_{3,15},\ S_{3,16},  \ S_{3,19}},
\end{equation}
\begin{equation}\label{change}
S_{3,2},\ S_{3,27}
\end{equation}
reproduces pp-waves that already appeared in \cite{php,hp} as duals with respect to some of the subgroups \eqref{restore}.
{Nevertheless, addition 
of suitable non-trivial $B$-field affects the dual model.
Contrary to \eqref{oneknown}, subgroups generated by \eqref{moreknown} produce different pp-wave backgrounds for different choices of initial torsionless $B$-field.} Moreover, subgroups generated by \eqref{change} give pp-waves only for particular non-vanishing $B$-field.

{Finally, introduction of $B$-fields yields new pp-waves obtained as T-duals of the flat torsionless
background by dualization with respect to 
subgroups generated by 
\begin{equation}\label{new}
S_{2,2},\ S_{2,12}.
\end{equation}
They are given by  the following metrics, torsions, dilatons, and
the Sfetsos--Tseytlin parameters.}
\begin{enumerate}
    \item The background with
    \begin{equation}\label{BrinkMetricS2,2}
 ds^{2}=\frac{2u^2 \left(u^4-5\right)}{\left(u^4+1\right)^2}\left(z_3^2+z_4^2\right)du^2 + 2 du dv +dz_3^2 + dz_4^2,
\end{equation}
$$ \wh {\cal H}=\frac{4 u}{u^4+1}\,du\wedge dz_3 \wedge dz_4,$$
$$ \Phi(u)=c_2+c_1 u+\log(u^4+1),$$
$$ g_1(u)=g_2(u)=u^2,\quad q=1 $$
{gained} by dualization with respect to the subgroup generated by $S_{2,2}$ and initial torsionless $B$-field with $B_6\neq 0$.
    \item {One-parametric set} of \bckg s with \begin{equation}\label{BrinkMetricS2,12}
\begin{array}{c}ds^{2}_\omega=\left[\frac{2 u^2 (u+1)^4-\omega^2 (2 u (5 u+4)+1)}{\left(u^2 (u+1)^2+\omega^2\right)^2}z_3^2+\frac{2 u^4 (u+1)^2-\omega^2 (2 u (5 u+6)+3)}{\left(u^2 (u+1)^2+\omega^2\right)^2}z_4^2\right]du^2\\+\ 2 du dv +dz_3^2 + dz_4^2,\end{array}
\end{equation}
$$ \wh {\cal H}_\omega=\frac{2 \omega (2 u+1)}{u^2 (u+1)^2+\omega^2} du\wedge dz_3 \wedge dz_4,$$
$$ \Phi_\omega(u)=c_2+c_1u+\log\left[u^2 (u+1)^2+\omega^2\right],$$
$$g_1(u')=(1+u)^2,\quad g_2(u)=u^2,\quad q=\omega$$
{gained} by dualization with respect to the subgroup generated by
$S_{2,12}$ and initial torsionless $B$-field with $B_6=\omega\neq0$.
\end{enumerate}


\section*{Acknowledgment}

This work was supported by the Grant Agency of the Czech Technical University in Prague, grant No. SGS16/239/OHK4/3T/14.

\begin{table}
\begin{center}
\footnotesize {\renewcommand{\arraystretch}{1.8}
\begin{tabular}{|c | c | c | c | }
\hline Metric and  torsion  in Brinkmann coordinates & Generated by  \\
\hline
\hline $\begin{array}{c}
   -\left(z_3^2+z_4^2\right)du^2 + 2 du dv +dz_3^2 + dz_4^2,
   \\2\,du\wedge dz_3 \wedge dz_4
 \end{array}$ &  $\begin{array}{c} S_{2,3}, S_{2,13},S_{2,21},S_{3,2},\\S_{3,27}, S_{4,17}, S_{4,29} \end{array}$ \\
\hline    $\begin{array}{c}
  2\frac{z_3^2}{u^2} du^2 + 2 du dv +dz_3^2 + dz_4^2,
\\0\end{array}$ &  $\begin{array}{c} S_{1,4},S_{2,8},S_{3,1},\\S_{3,4},S_{3,8},S_{3,15}\end{array}$ \\
\hline    $\begin{array}{c}
  2\frac{z_3^2+z_4^2}{u^2} du^2 + 2 du dv +dz_3^2 + dz_4^2,
\\0\end{array}$ &  $ S_{2,2}$\\
\hline
$\begin{array}{c}2\left(\frac{{z_3}^2}{u^2}+\frac{{z_4}^2}{(1+u)^2}\right)du^2+2dudv+d{z_3}^2+d{z_4}^2\\0\end{array}$ & $S_{2,12}$ \\
\hline $\begin{array}{c}
  \frac{ \left(2 u^2-1\right)z_3^2-3 z_4^2}{\left(u^2+1\right)^2}du^2+2dudv+dz_3^2+dz_4^2,
\\  \frac{2}{u^2+1}\,du\wedge dz_3 \wedge dz_4
 \end{array}$ & $\begin{array}{c} S_{1,13},S_{2,8},S_{3,4},S_{3,8},S_{3,15},S_{3,16},\\S_{3,17},S_{3,19},S_{3,21},S_{4,23},{S_{4,25}}\end{array}$ \\
\hline $\begin{array}{c}
      2\frac{z_3^2}{\sinh^2 u} du^2 + 2 du dv +dz_3^2 + dz_4^2,\\ 0 \end{array}$ & $ \begin{array}{c} S_{3,22},S_{3,23},S_{3,25},\\ S_{3,26},S_{4,7},S_{4,8}\end{array}$ \\
\hline $\begin{array}{c}
      -2\frac{z_3^2}{\cosh^2 u} du^2 + 2 du dv +dz_3^2 + dz_4^2,\\ 0 \end{array}$ & $ \begin{array}{c}S_{3,22},S_{3,23},S_{3,25},\\ S_{3,26},S_{4,7},S_{4,8}\end{array}$ \\
\hline $\begin{array}{c}
      2\frac{z_3^2+z_4^2}{\sinh^2 u} du^2 + 2 du dv +dz_3^2 + dz_4^2,\\ 0 \end{array}$ & $ \begin{array}{c}S_{4,26},S_{4,27}\end{array}$ \\
\hline $\begin{array}{c}
      -2\frac{z_3^2+z_4^2}{\cosh^2 u} du^2 + 2 du dv +dz_3^2 + dz_4^2,\\ 0 \end{array}$ & $ \begin{array}{c}S_{4,26},S_{4,27}\end{array}$ \\
\hline $\begin{array}{c}
-\left(\frac{1\pm 2 \tan^2{\gamma}\, \text{sech}^2(u)}{\tan^2{\gamma}}\right)\left(z_3^2+z_4^2\right) du^2 + 2 du dv +dz_3^2 + dz_4^2,
\\
 \frac{2}{\tan{\gamma}} du\wedge dz_3 \wedge dz_4
 \end{array}$ & $\begin{array}{c}S_{4,28}\\0< \gamma < \pi, \gamma \neq \frac{\pi}{2}\end{array}$ \\
\hline $\begin{array}{c}
 -\left(\frac{4 \left(2 \beta ^2\mp\left(\beta ^2+1\right) \cosh (2 u)+1\right)}{\left(\beta ^2\pm\left(\beta ^2+1\right) \cosh (2 u)-1\right)^2} z_3^2 +\frac{4 \beta ^2 \left(\beta ^2\pm\left(\beta ^2+1\right) \cosh (2 u)+2\right)}{\left(\beta ^2\pm\left(\beta ^2+1\right) \cosh (2 u)-1\right)^2} z_4^2\right) du^2$$ $$ \\ + 2 du dv +dz_3^2 + dz_4^2,\\ \frac{4 \beta }{\pm \beta ^2 + \left(\beta ^2+1\right) \cosh (2 u)\mp 1} du\wedge dz_3 \wedge dz_4 \end{array}$ &  $\begin{array}{c}S_{4,31}, S_{4,33}\\ \beta\neq0\end{array}$ \\
\hline
 $\begin{array}{c}
   \frac{2u^2 \left(u^4-5\right)}{\left(u^4+1\right)^2}\left(z_3^2+z_4^2\right)du^2 + 2 du dv +dz_3^2 + dz_4^2,
\\
   \frac{4 u}{u^4+1}\,du\wedge dz_3 \wedge dz_4
 \end{array}$ &  $S_{2,2}$ \\

\hline $\begin{array}{c} \left[\frac{2 u^2 (u+1)^4-\omega^2 (2 u (5 u+4)+1)}{\left(u^2 (u+1)^2+\omega^2\right)^2}z_3^2+\frac{2 u^4 (u+1)^2-\omega^2 (2 u (5 u+6)+3)}{\left(u^2 (u+1)^2+\omega^2\right)^2}z_4^2\right]du^2 \\ + 2 du dv +dz_3^2 + dz_4^2, \\ \frac{2 \omega (2 u+1)}{u^2 (u+1)^2+\omega^2} du\wedge dz_3 \wedge dz_4  \end{array}$ &  $\begin{array}{c}S_{2,12}\\\omega\neq0\end{array}$ \\

\hline
\end{tabular}
\normalsize \caption{{Complete list of pp-waves obtainable as (non-)Abelian T-duals of the flat metric with torsionless $B$-field}\label{tableconcl1} }
}\end{center}
\end{table}

\appendix
\section*{Appendix}
\addcontentsline{toc}{section}{Appendix}

Below, we summarize dual backgrounds {with pp-wave metrics} produced by dualization with initial torsionless $B$-fields that{, however,} already appeared in \cite{php,hp} where the results were obtained without initial $B$-field. 
{In case of one-dimensional subgroups, influence of the \tbf{}  on the dual backgrounds can be completely eliminated by coordinate and gauge \tfn s {in general}. Therefore, we omit these subgroups also from the Appendix and the results can be found in \cite{php}.}

\subsection*{A.1\mez Duals with respect to two-dimensional subgroups}\label{app2dim}

As mentioned in the beginning of  Sec. \ref{2dimsubgps}, we can eliminate $B_1,\dots,B_5$ and the only relevant component is the constant $B_6$.

\subsubsection*{Dualization with respect to Abelian subgroup generated by $S_{2,8}$}

We choose $\xi(s) = (-\frac{s_2}{2s_1},0,0,s_1+\frac{s_2}{2s_1})$. After elimination of $B_1,\dots,B_5$, we get the dual background in the form
$$\wh\cf(s,\hat g)=\left(
\begin{array}{cccc}
 1-\frac{s_2}{s_1^2} & \frac{1}{2
   s_1} & 0 & 0 \\
 \frac{1}{2 s_1} & 0 & 0 & 0 \\
 0 & 0 & \frac{1}{B_6^2+s_1^2}
   &-\frac{B_6}{B_6^2+s_1^2} \\
 0 & 0 & \frac{B_6}{B_6^2+s_1^2} &
   \frac{s_1^2}{B_6^2+s_1^2}\\
\end{array}
\right).$$
The coordinate transformation
\begin{align*}
s_1&=u, & s_2 &=-u^2+2 u v+ \frac{B_6^2 z_4^2-u^2 z_3^2}{B_6^2+u^2},\\ s_3&=z_3 \sqrt{B_6^2+u^2}, & \hat g_1&=\frac{z_4 \sqrt{B_6^2+u^2}}{u}
\end{align*}
{brings the dual metric to the Brinkmann form}
$$ds^2=\frac{(2 u^2-B_6^2)z_3^2-3 B_6^2 z_4^2}{\left(u^2+B_6^2\right)^2}du^2 + 2 du dv +dz_3^2 + dz_4^2,$$
{while the torsion reads}
$$
\wh {\cal H}=\frac{2B_6}{u^2+B_6^2} du\wedge dz_3 \wedge dz_4.
$$
This restores the result of \cite{php} for $B_6=0$. {In the particular case of} $B_6\neq0$, the constant $B_6$ can be effectively rescaled to $1$ by subsequent coordinate transformation, resulting in {another} metric and torsion that appeared already in \cite{php,hp}, see Tab. \ref{tableconcl1}.

\subsubsection*{Dualization with respect to Abelian subgroup generated by $S_{2,13}$}

We choose $\xi(s) = (0,s_2,s_1,0)$. After elimination of $B_1,\dots,B_5$, we get the dual background in the form
$$\wh\cf(s,\hat g)= \left(
\begin{array}{cccc}
 1 & 0 & 0 &
   -\frac{s_2}{B_6+2 {\epsilon}} \\
 0& 1 & 0 &
   \frac{s_1}{B_6+2 {\epsilon}} \\
 0 & 0 & 0 & -\frac{1}{B_6+2 {\epsilon}} \\
 -\frac{s_2}{B_6-2 {\epsilon}} &
   \frac{s_1}{B_6-2 {\epsilon}} &
   \frac{1}{B_6-2 {\epsilon}} &
   \frac{s_1^2+s_2^2}{B_6^2-4
  } \\
\end{array}
\right)$$
where $\epsilon = \pm 1$. Subsequent coordinate transformation
\begin{align*}
s_1 &=z_3 \sin \frac{B_6 u}{2}+z_4 \cos \frac{B_6 u}{2}, & \hat g_1 &=\epsilon\,s_2, \\ s_2 &=z_3 \cos \frac{B_6 u}{2}-z_4 \sin \frac{B_6u}{2}, & \hat g_2 &=\frac{1}{2} \left(B_6^2-4\right) u
\end{align*}
brings the metric and torsion to the forms \eqref{hom-isotr-pp} and \eqref{hom-isotr-H}.

\subsubsection*{Dualization with respect to non-Abelian subgroup generated by $S_{2,21}$}
We choose $\xi(s) = (\frac{1}{2},s_2,-s_1,\frac{1}{2})$. After elimination of $B_1,\dots,B_5$, the dual background acquires the form
$$\wh\cf(s,\hat g)= \left(
\begin{array}{cccc}
 1 & 0 & 0 & \frac{s_2 \cos \gamma}{B_6+(\hat g_2-1) \sin \gamma} \\
 0 & 1 & 0 & -\frac{s_1 \cos \gamma}{B_6+(\hat g_2-1) \sin \gamma} \\
 0 & 0 & 0 & \frac{1}{-B_6-\hat g_2 \sin \gamma+\sin \gamma} \\
 \frac{s_2 \cos \gamma}{B_6+(\hat g_2+1) \sin \gamma} & -\frac{s_1 \cos \gamma}{B_6+(\hat g_2+1) \sin
   \gamma} & \frac{1}{B_6+(\hat g_2+1) \sin \gamma} & \frac{\left(s_1^2+s_2^2\right) \cos
   ^2\gamma}{B_6^2+2 \hat g_2 \sin \gamma B_6+\left(\hat g_2^2-1\right) \sin ^2\gamma} \\
\end{array}
\right) $$
where $0 < \gamma < \pi, \gamma \neq \frac{\pi}{2}$. After the coordinate transformation
$$s_1=z_3 \cos \left[\cot \gamma \log (\cosh (u\,\tan \gamma))\right]+z_4 \sin \left[\cot \gamma \log (\cosh (u\,\tan \gamma))\right],$$
\begin{equation}\label{tfn221}
s_2=z_4 \cos \left[\cot \gamma \log (\cosh (u\,\tan \gamma))\right]-z_3 \sin \left[\cot\gamma \log (\cosh (u\,\tan \gamma))\right],
   \end{equation}
   $$\ghat_1={v} \cos \gamma,\qquad \ghat_2=\tanh (u\,\tan \gamma)-\frac{1}{\sin\gamma} B_{6}$$
valid for $|\ghat_2+\frac{1}{\sin\gamma} B_{6}|<1$,  the background  gives the metric \eqref{hom-isotr-pp} and torsion \eqref{hom-isotr-H}. Changing $\cosh (u\,\tan \gamma) \rightarrow \sinh (u\,\tan \gamma)$, $\tanh (u\,\tan \gamma)\rightarrow \coth (u\,\tan \gamma)$ in (\ref{tfn221}), we get the transformation for $|\ghat_2+\frac{1}{\sin\gamma} B_{6}|>1$.

\subsection*{A.2\mez Duals with respect to Abelian three-dimensional subgroups}\label{3dimnoB}

Abelian subgroups are generated by $S_{3,1},\ldots,S_{3,9}$. {Based on the conclusions of} Sec. \ref{elimab}, we find that {the only relevant \comp s of the torsionless background are}
$$B_4(s_1)=B_4=const.,\qquad B_5(s_1)=B_5=const.,\qquad B_6(s_1)=B_6=const.$$

\subsubsection*{Dualization with respect to subgroup generated by $S_{3,1}$}

This is one of the cases where dualization can be performed only for non-vanishing $B$-field, namely for $B_5^2+B_6^2\neq0$. We choose $\xi(s)=(s_1,0,0,0)$,  $s_1\neq 0$.
After elimination of $B_1$, $B_2$, and $B_3$, 
we get the dual background with vanishing scalar curvature $\wh{\cal R}$ and torsion $\wh{\cal H}$ in the form
$$\wh{\cal F}(s,\hat g)=\left(
\begin{array}{cccc}
 -\frac{B_4^2+u^2 \left(B_5^2+B_6^2+u^2\right)}{\left(B_5^2+B_6^2\right) u^2} & \frac{B_5 u^2+B_4
   B_6}{\left(B_5^2+B_6^2\right) u^2} & \frac{B_6 u^2-B_4 B_5}{\left(B_5^2+B_6^2\right) u^2} &
   \frac{u^4+B_4^2}{\left(B_5^2+B_6^2\right) u^2} \\
 \frac{B_5 u^2-B_4 B_6}{\left(B_5^2+B_6^2\right) u^2} & \frac{B_6^2}{\left(B_5^2+B_6^2\right) u^2} & -\frac{B_5
   B_6}{\left(B_5^2+B_6^2\right) u^2} & \frac{B_4 B_6-B_5 u^2}{\left(B_5^2+B_6^2\right) u^2} \\
 \frac{B_6 u^2+B_4 B_5}{\left(B_5^2+B_6^2\right) u^2} & -\frac{B_5 B_6}{\left(B_5^2+B_6^2\right) u^2} &
   \frac{B_5^2}{\left(B_5^2+B_6^2\right) u^2} & -\frac{B_6 u^2+B_4 B_5}{\left(B_5^2+B_6^2\right) u^2} \\
 -\frac{u^4+B_4^2}{\left(B_5^2+B_6^2\right) u^2} & \frac{B_5 u^2+B_4 B_6}{\left(B_5^2+B_6^2\right) u^2} & \frac{B_6
   u^2-B_4 B_5}{\left(B_5^2+B_6^2\right) u^2} & \frac{u^4+B_4^2}{\left(B_5^2+B_6^2\right) u^2} \\
\end{array}
\right).$$
Using the transformation of coordinates
\begin{align*}
s_1&=u,\\
\hat{g}_1 &=\frac{B_5 \left(-\frac{B_4^2}{u}+B_5^2 u+B_6^2 u+\frac{u^3}{3}\right)}{2 \left(B_5^2+B_6^2\right)}+\frac{B_6\left(x_1-B_4 x_2\right)}{\sqrt{B_5^2+B_6^2}}+B_5 v,\\
\hat{g}_2 &=\frac{B_6 \left(-\frac{B_4^2}{u}+B_5^2 u+B_6^2 u+\frac{u^3}{3}\right)}{2 \left(B_5^2+B_6^2\right)}+\frac{B_5 \left(B_4 x_2-x_1\right)}{\sqrt{B_5^2+B_6^2}}+B_6 v,\\
\hat{g}_3&=x_2 \sqrt{B_5^2+B_6^2},
\end{align*}
the metric acquires the Rosen form\footnote{As mentioned in the Introduction, the Rosen form of metric is not unique in general. This metric can be also found in the Rosen form $ds^2=2dudv+\frac{1}{u^2}dx_1^2+dx_2^2$ corresponding to the Sfetsos--Tseytlin parameters $(1,u^2,0)$.}
$$ds^2=2dudv+\frac{1}{u^2}dx_1^2+u^2dx_2^2$$
corresponding to the Sfetsos--Tseytlin parameters $\left(\frac{1}{u^2},u^2,0\right)$. The dual torsion vanishes.
Consequently, it can be expressed via \eqref{BrinkmannRosen1}  and
\eqref{BrinkmannRosen} in the Brinkmann form
\begin{equation}\label{ppwS3,1} ds^2=2\frac{z_3^2}{u^2} du^2 + 2 du
dv +dz_3^2 + dz_4^2
\end{equation}
that appeared already in \cite{php}.

\subsubsection*{Dualization with respect to subgroup generated by $S_{3,4}$}

We choose $\xi(s)=(s_1,0,0,0)$ with $B_4^2+B_6^2\,s_1^2\neq0$, $s_1\neq 0$.
After elimination of $B_1$, $B_2$, and $B_3$, 
we get the dual background with vanishing scalar curvature $\wh{\cal R}$ in the form
$$\wh{\cal F}(s,\hat g)=\left(
\begin{array}{cccc}
 -\frac{B_4^2+B_5^2+\left(B_6^2+1\right) u^2}{B_4^2+B_6^2 u^2} & \frac{B_4-B_5 B_6}{B_4^2+B_6^2 u^2} &
   \frac{B_5^2+u^2}{B_4^2+B_6^2 u^2} & -\frac{B_6 u^2+B_4 B_5}{B_4^2+B_6^2 u^2} \\
 \frac{B_4+B_5 B_6}{B_4^2+B_6^2 u^2} & \frac{B_6^2}{B_4^2+B_6^2 u^2} & -\frac{B_4+B_5
   B_6}{B_4^2+B_6^2 u^2} & \frac{B_4 B_6}{B_4^2+B_6^2 u^2} \\
 -\frac{B_5^2+u^2}{B_4^2+B_6^2 u^2} & \frac{B_4-B_5 B_6}{B_4^2+B_6^2 u^2} & \frac{B_5^2+u^2}{B_4^2+B_6^2
   u^2} & -\frac{B_6 u^2+B_4 B_5}{B_4^2+B_6^2 u^2} \\
 \frac{B_4 B_5-B_6 u^2}{B_4^2+B_6^2 u^2} & \frac{B_4 B_6}{B_4^2+B_6^2 u^2} & \frac{B_6 u^2-B_4
   B_5}{B_4^2+B_6^2 u^2} & \frac{B_4^2}{B_4^2+B_6^2 u^2} \\
\end{array}
\right).$$
Dual torsion vanishes for $B_4=0$ or $B_6=0$. Using the transformation of coordinates
\begin{align*}
s_1 &=u,\\
\hat{g}_1 & =\frac{u \left(3 B_4^2+3 B_5^2+u^2\right)}{6 B_4}+B_4 v,\\
\hat{g}_2 & =\frac{B_5 B_6 u}{B_4}+x_2,\\
\hat{g}_3 &=\frac{B_6 \left(-3 B_4^2\left(u+2v\right)+3 B_5^2 u+u^3\right)+6 B_4 B_5 x_2+6 B_4 x_1}{6 B_4^2},
\end{align*}
valid for\footnote{There is a \tfn{} for $B_4$=0 as well, giving the following results with $B_4=0$.} $B_4\neq 0$, the metric acquires the Rosen form
$$ds^2=2dudv+\frac{1}{B_4^2+B_6^2u^2}dx_1^2+\frac{u^2}{B_4^2+B_6^2u^2}dx_2^2.$$
Consequently, it can be expressed via \eqref{BrinkmannRosen1} and \eqref{BrinkmannRosen} in the Brinkmann coordinates as
\begin{equation}
ds^2=\frac{B_6^2 \left( \left(2 B_6^2 u^2-B_4^2\right)z_3^2-3 B_4^2 z_4^2\right)}{\left(B_4^2+B_6^2 u^2\right)^2}du^2+2dudv+dz_3^2+dz_4^2
\label{General1BrinkmannS34}
\end{equation}
{ with the corresponding torsion}
$$\wh {\cal H}=\frac{2 B_4 B_6}{B_4^2+B_6^2u^2} du\wedge dz_3 \wedge dz_4.
$$
For $B_6=0$, the dual background is flat and torsionless. {In the particular case of} $B_6\neq 0$, the Brinkmann metric (\ref{General1BrinkmannS34}) can be transformed
to the form
$$ds^2=\frac{ \left(2 u^2-B_4^2\right)z_3^2-3 B_4^2 z_4^2}{\left(u^2+B_4^2\right)^2}du^2+2dudv+dz_3^2+dz_4^2$$
where effectively $B_4\in\{0,1\}$ due to additional coordinate transformation. All these backgrounds appeared already in \cite{php,hp}.

\subsubsection*{Dualization with respect to subgroup generated by $S_{3,8}$}

We choose $\xi(s)=(s_1,0,0,0)$ with $B_5^2+B_6^2\neq0$, $s_1\neq 0$, $s_1\neq -1$.
After elimination of $B_1$, $B_2$, and $B_3$, 
we get the dual background with vanishing scalar curvature $\wh{\cal R}$ in the form
\begin{align*}
&\wh {\cal F}(s, \hat g)=\\
&\left(
\begin{array}{cccc}
 -\frac{B_4^2+B_5^2 (s_1+1)^2+s_1^2 \left(B_6^2+(s_1+1)^2\right)}{B_6^2 s_1^2+B_5^2 (s_1+1)^2} & \frac{B_5 (s_1+1)^2+B_4 B_6}{B_6^2
   s_1^2+B_5^2 (s_1+1)^2} & \frac{B_6 s_1^2-B_4 B_5}{B_6^2 s_1^2+B_5^2 (s_1+1)^2} & \frac{B_4^2+s_1^2 (s_1+1)^2}{B_6^2 s_1^2+B_5^2
   (s_1+1)^2} \\
 \frac{B_5 (s_1+1)^2-B_4 B_6}{B_6^2 s_1^2+B_5^2 (s_1+1)^2} & \frac{B_6^2}{B_6^2 s_1^2+B_5^2 (s_1+1)^2} & -\frac{B_5
   B_6}{B_6^2 s_1^2+B_5^2 (s_1+1)^2} & \frac{B_4 B_6-B_5 (s_1+1)^2}{B_6^2 s_1^2+B_5^2 (s_1+1)^2} \\
 \frac{B_6 s_1^2+B_4 B_5}{B_6^2 s_1^2+B_5^2 (s_1+1)^2} & -\frac{B_5 B_6}{B_6^2 s_1^2+B_5^2 (s_1+1)^2} &
   \frac{B_5^2}{B_6^2 s_1^2+B_5^2 (s_1+1)^2} & -\frac{B_6 s_1^2+B_4 B_5}{B_6^2 s_1^2+B_5^2 (s_1+1)^2} \\
 -\frac{B_4^2+s_1^2 (s_1+1)^2}{B_6^2 s_1^2+B_5^2 (s_1+1)^2} & \frac{B_5 (s_1+1)^2+B_4 B_6}{B_6^2 s_1^2+B_5^2 (s_1+1)^2} & \frac{B_6
   s_1^2-B_4 B_5}{B_6^2 s_1^2+B_5^2 (s_1+1)^2} & \frac{B_4^2+s_1^2 (s_1+1)^2}{B_6^2 s_1^2+B_5^2 (s_1+1)^2} \\
\end{array}
\right).
\end{align*}
Dual torsion vanishes for $B_5=0$ or $B_6=0$. Using the transformation of coordinates 
\begin{align*}
s_1 &=u,\\
\hat{g}_1 &=\frac{-3 B_4^2+(u+1) \left(3 B_5^2 (u+2 v)+u^3\right)}{6 B_5 (u+1)},\\
\hat{g}_2 &=\frac{B_4^2 B_6}{2 B_5^2 (u+1)}+\frac{B_4 x_2}{B_5}-\frac{B_6
   u^3}{6 B_5^2}+\frac{x_1}{B_5}+\frac{ B_6 u}{2}+B_6 v,\\
\hat{g}_3 &=\frac{B_4
   B_6}{B_5 u+B_5}+x_2,
\end{align*}
valid for\footnote{There is a \tfn{} for $B_5$=0 as well, giving the following results with $B_5=0$.} $B_5\neq 0$ the metric acquires the Rosen form
$$ds^2=2dudv+\frac{1}{B_5^2(1+u)^2+B_6^2u^2}dx_1^2+\frac{u^2(1+u)^2}{B_5^2(1+u)^2+B_6^2u^2}dx_2^2.$$
Consequently, it can be expressed via \eqref{BrinkmannRosen1} and \eqref{BrinkmannRosen} in the Brinkmann coordinates as
\begin{equation}
ds^2=\frac{\left(2 B_5^4 (u+1)^2+B_5^2 B_6^2 \left(4 u^2+4 u-1\right)+2 B_6^4 u^2\right)z_3^2-3 B_5^2 B_6^2 z_4^2}{\left(B_5^2(u+1)^2+B_6^2 u^2\right)^2}du^2+2dudv+dz_3^2+dz_4^2
\label{General1BrinkmannS38}
\end{equation}
 with the corresponding torsion
$$\wh {\cal H}=\frac{2 B_5 B_6}{B_5^2(1+u)^2+B_6^2u^2} du\wedge dz_3 \wedge dz_4.
$$
For $B_5=0$ or $B_6=0$, the Brinkmann metric
(\ref{General1BrinkmannS38}) acquires the form
\eqref{ppwS3,1} and the torsion vanishes. {In the particular case of} $B_5\neq 0$ and $B_6\neq 0$, the Brinkmann metric (\ref{General1BrinkmannS38}) can be transformed to the form
 \begin{equation}\label{ppwS3,17} ds^2=\frac{
\left(2 u^2-1\right)z_3^2-3
z_4^2}{\left(u^2+1\right)^2}du^2+2dudv+dz_3^2+dz_4^2
\end{equation}with torsion \begin{equation}\label{torS3,17}
\wh {\cal H}=\frac{2}{u^2+1} du\wedge dz_3 \wedge dz_4.
\end{equation} All these
backgrounds appeared already in \cite{php,hp}.

\subsection*{A.3\mez Duals with respect to Heisenberg three-dimensional subgroups}
Heisenberg subgroups producing pp-waves as duals of flat torsionless background are generated by $S_{3,15}$, $S_{3,16}$, $S_{3,17}$, $S_{3,19}$, and $S_{3,21}$. In these cases, a torsionless background on $M$ is gained if the matrix $B(s_1)$ fulfills
$$B_3(s_1)=\frac{1}{K} B_4'(s_1),\qquad B_5(s_1)=B_5=const.,\qquad B_6(s_1)=B_6=const.$$
where $K=\beta\neq 0$ for $S_{3,15}$, $K=-2\epsilon=\mp 2$ for $S_{3,16},
S_{3,17}$, $K=1$ for $S_{3,19}$, and   $K=-\beta\neq 0$ for $S_{3,21}.$
Dependence of these dual models on $B_1(s_1)$,
$B_2(s_1)$, $B_3(s_1)$, and $B_4(s_1)$ can be eliminated by the
coordinate \tfn{}
\begin{equation}\label{eliminate3B1H}\hat{g}_1\mapsto \hat g_1+\int^{s_1} B_1(r) \, dr,\qquad \hat{g}_2\mapsto \hat g_2+\int^{s_1} B_2(r) \, dr,\qquad \hat{g}_3\mapsto \hat g_3+\frac{1}{K}B_4(s_1)
\end{equation}
leaving the dual \bckg{} dependent only on constants $B_5$ and $B_6$.

\subsubsection*{Dualization with respect to subgroup generated by $S_{3,15}$}

Dualization can be performed only for $B_5^2+B_6^2\neq 0$. We choose
$\xi(s)=(s_1,0,0,0)$ with $(1 + \beta^2) B_5^2 + 2 \beta B_5 B_6 s_1
+ B_6^2 s_1^2\neq0$, $s_1 \neq0$.  After elimination of $B_1$,
$B_2$, $B_3$, and $B_4$ by the coordinate \tfn{}
\eqref{eliminate3B1H}, the dual background $\wh \cf(s,\hat g)$ still has rather complicated { form. Nevertheless, the Ricci tensor has only one non-vanishing component
$$\wh{\cal R}_{s_1s_1}(s,\hat{g})=\frac{2 B_6^2 \left(\left(\beta^2-2\right) B_5^2+2 \beta B_5 B_6 s_1+ B_6^2 s_1^2\right)}{\left(\left(\beta^2+1\right) B_5^2+2 \beta B_5 B_6 s_1+B_6^2s_1^2\right){}^2}$$ and scalar curvature $\wh {\cal R}$ vanishes}. It is pp-wave background with covariantly constant null vector field $k=B_5\partial_{\hat{g}_1}+B_6\partial_{\hat{g}_2}$.
For $B_6=0$,
the dual background is flat and torsionless. {In the particular case of} $B_5=0$, {the dual background reads
$$\wh {\cal F}(s, \hat g)=\left(
\begin{array}{cccc}
 -\frac{{\hat g_3}^2 \beta+
   s_1^2}{B_6^2s_1^2}-1 &
   \frac{({s_1}-{\hat g_3}) \beta }{{B_6}
   s_1^2} & \frac{1}{B_6} &
   \frac{s_1^2+{\hat g_3}^2 \beta
   ^2}{B_6^2 s_1^2} \\
 \frac{({\hat g_3}+{s_1}) \beta }{{B_6}
   s_1^2} & \frac{1}{s_1^2} & 0 &
   -\frac{({\hat g_3}+{s_1}) \beta }{{B_6}
   s_1^2} \\
 \frac{1}{B_6} & 0 & 0 & -\frac{1}{B_6}
   \\
 -\frac{s_1^2+{\hat g_3}^2 \beta
   ^2}{B_6^2 s_1^2} &
   \frac{({s_1}-{\hat g_3}) \beta }{{B_6}
   s_1^2} & \frac{1}{B_6} &
   \frac{s_1^2+{\hat g_3}^2 \beta
   ^2}{B_6^2 s_1^2} \\
\end{array}
\right).$$
Using} the transformation of coordinates
\begin{align*}
s_1&=u,\\
\hat{g}_1&=-\frac{\beta u^2}{2 B_6}+\frac{1}{2} \beta B_6 z_4^2+uz_3,\\
\hat{g}_2&=\frac{u^2 \left(\beta^2+B_6^2+1\right)+B_6^2 \left(-\beta^2 z_4^2+2 u v-z_3^2\right)}{2 B_6 u},\\
\hat{g}_3&= B_6 z_4,
\end{align*}
the metric acquires the Brinkmann form \eqref{ppwS3,1} and torsion vanishes. {In the case of} $B_6\neq 0$ and $B_5\neq0$, using rather complicated  transformation of coordinates, the metric acquires the Brinkmann form \eqref{ppwS3,17} with torsion
\eqref{torS3,17}.

\subsubsection*{Dualization with respect to subgroups generated by $S_{3,16},S_{3,17}$}
\label{S316317} 
{The algebra $S_{3,17}$ is equal
to $S_{3,16}$ for $\beta=0$. Dualization with respect to $
S_{3,16},S_{3,17}$ can be performed only for non-trivial $B$-field.
We choose $\xi(s)=(s_1,0,0,0)$  with
$$1+\beta s_1+s_1^2\neq 0,\ \beta^2 B_5^2 + 2 \beta B_5 B_6 \epsilon + (B_5^2 + B_6^2) + 2 \beta B_5^2 s_1 + (B_5^2 + B_6^2)s_1^2 \neq 0.$$
After elimination of $B_1$, $B_2$, $B_3$, and $B_4$ by the
coordinate \tfn{} \eqref{eliminate3B1H}, the dual background
$\wh \cf(s,\hat g)$ still has rather complicated form. 
Nevertheless, it has simple Ricci tensor with only one non-vanishing component
\begin{align*}
&\wh{\cal R}_{s_1s_1}(s,\hat{g})=\\
&=\frac{2 \left(\beta^2 \left(B_5^4-2 B_5^2 B_6^2\right)+2 \beta B_5 \left(B_5^2+B_6^2\right) (B_5 s_1-2 B_6 \epsilon)-\left(2-s_1^2\right) \left(B_5^2+B_6^2\right)^2\right)}{\left(\beta^2 B_5^2+2 \beta B_5 (B_5s_1+B_6 \epsilon)+\left(s_1^2+1\right)\left(B_5^2+B_6^2\right)\right)^2}
\end{align*}
and vanishing scalar curvature $\wh {\cal R}$. It is a pp-wave background with covariantly constant null vector field $k=B_5\partial_{\hat{g}_1}+B_6\partial_{\hat{g}_2}$.
For $B_5\neq 0$, $B_6\neq 0$, and $\beta=-\varepsilon \left(\frac{B_5}{B_6}+\frac{B_6}{B_5}\right)$, using rather complicated coordinate transformation, the metric acquires the Brinkmann form \eqref{ppwS3,1} and the torsion vanishes. In case of $B_5\neq 0$, $B_6\neq 0$, and $\beta\neq-\varepsilon \left(\frac{B_5}{B_6}+\frac{B_6}{B_5}\right)$ or $B_5=0$ or $B_6=0$  or $\beta=0$ ($S_{3,17}$), the metric and torsion can be brought to the forms \eqref{ppwS3,17} and \eqref{torS3,17}.}

\subsubsection*{Dualization with respect to subgroup generated by $S_{3,19}$}

We choose $\xi(s)=(s_1,0,0,0)$ with $B_5^2 + 2 B_5 B_6 s_1 + B_6^2 (1+ s_1^2) \neq 0 $. After elimination of $B_1$, $B_2$, $B_3$, and $B_4$ by the coordinate \tfn{} \eqref{eliminate3B1H}, the dual background $\wh \cf(s,\hat g)$ still has rather complicated form, nevertheless, it has simple Ricci tensor with only one non-vanishing component
$$\wh{\cal R}_{s_1s_1}(s,\hat{g})=\frac{2 B_6^2 \left(B_5^2+2B_5 B_6 s_1+B_6^2\left(-2+
s_1^2\right)\right)}{B_5^2+ 2 B_5 B_6 s_1 + B_6^2\left(1+ s_1^2\right)}$$
and vanishing scalar curvature $\wh {\cal R}$. It is pp-wave background with covariantly constant null vector field $k=B_5\partial_{\hat{g}_1}+B_6\partial_{\hat{g}_2}$.
For $ B_6=0 $, the dual background is flat and  torsionless. {In the particular case of} $B_5=0$, the metric can be brought to the Brinkmann form \eqref{ppwS3,1} and torsion vanishes. {In the particular case of} $B_5\neq 0$ and  $B_6\neq 0$, the dual metric acquires the Brinkmann form \eqref{ppwS3,17} with torsion \eqref{torS3,17} after the coordinate \tfn
\begin{align*}
s_1=&u-\frac{B_5}{B_6},\\
\hat{g}_1=&\frac{B_5 \left(u^3 \left(1-B_6^2 \left(z_4^2-1\right)\right)+2 B_6^2 u^2 v+u \left(1-B_6^2 \left(z_3^2+2 z_4^2-1\right)\right)+2 B_6^2 v\right)}{2 B_6^2 \left(u^2+1\right)}+\\ & \frac{u^2 \left(B_6^2 z_4^2-1\right)+B_6^2 z_4^2}{2 B_6}+\sqrt{1+u^2} z_3,\\
\hat{g}_2 =&\frac{u^3 \left(1-B_6^2 \left(z_4^2-1\right)\right)+2 B_6^2 u^2 v+u \left(1-B_6^2 \left(z_3^2+2 z_4^2-1\right)\right)+2 B_6^2 v}{2 B_6 \left(u^2+1\right)},\\
\hat{g}_3=&B_6\sqrt{1+u^2}\, {z_4}.
\end{align*}

\subsubsection*{Dualization with respect to subgroup generated by $S_{3,21}$}

We choose $\xi(s)=(s_1,0,0,0)$ with $(1 + \beta^2) B_5^2 + 2 B_5 B_6 \epsilon + B_6^2 \epsilon^2 - 2 \beta B_5 B_6 s_1 +
 B_6^2 s_1^2 \neq0$, $s_1+\beta \epsilon \neq0$. After elimination of $B_1$, $B_2$, $B_3$, and $B_4$ by the coordinate \tfn{} \eqref{eliminate3B1H}, the dual background $\wh \cf(s,\hat g)$ still has rather complicated form, nevertheless, it has simple Ricci tensor with only one non-vanishing component
$$\wh{\cal R}_{s_1s_1}(s,\hat{g})=\frac{2 B_6^2 \left(\left(\beta^2-2\right) B_5^2-2 \beta B_5 B_6 s_1-4 B_5 B_6 \epsilon-2 B_6^2 \epsilon^2+B_6^2
   s_1^2\right)}{\left(\left(\beta^2+1\right) B_5^2-2 \beta B_5 B_6 s_1+2 B_5 B_6 \epsilon+B_6^2 \epsilon^2+B_6^2 s_1^2\right){}^2}$$
and vanishing scalar curvature $\wh {\cal R}$. It is pp-wave background with covariantly constant null vector field $k=B_5\partial_{\hat{g}_1}+B_6\partial_{\hat{g}_2}$.
For $ B_6=0 $, the dual background is flat and torsionless.
{In the particular case of} $B_6\neq0$, the metric and torsion can be brought to the forms  \eqref{ppwS3,17} and \eqref{torS3,17}.

\subsection*{A.4\mez Duals with respect to subgroups generated by Bianchi $4$, Bianchi $5$}\label{bia45}

\subsubsection*{Dualization with respect to subgroups generated by $S_{3,22},S_{3,25}$}

In these cases, a torsionless background on $M$ is obtained if the matrix $B(s_1)$ fulfills
$$B_2(s_1)=- B_4'(s_1)+\alpha\,B_5'(s_1),\qquad B_3(s_1)=- B_5'(s_1),\qquad B_6(s_1)=0$$
where $\alpha=0$ for $S_{3,25}$ and $\alpha>0$ for $S_{3,22}$. We choose $\xi(s)=(\half,0,s_1,\half)$. The sigma model is dualizable for $B_5(s_1)\neq \pm1$. The dual background $\wh\cf(s,\hat g)$ has rather complicated form with vanishing scalar curvature $\wh {\cal R}$ and torsion $\wh {\cal H}$. Dependence of the dual model on components of the matrix $B(s_1)$ can be eliminated by the \coor{} \tfn
\begin{equation*}
\hat{g}_1\mapsto\hat{g}_1+ \int^{s_1} B_1(r) \, dr, \qquad
\hat{g}_2\mapsto\hat{g}_2-B_4(s_1)+\alpha\,B_5(s_1),\qquad
\hat{g}_3\mapsto\hat{g}_3-B_5(s_1)
\end{equation*}
that brings the dual background to the form
$$\wh\cf(s,\hat g)=\left(
\begin{array}{cccc}
 1 & 0 & 0 & 0 \\
 0 & 0 & 0 & \frac{1}{1-\hat{g}_3} \\
 0 & 0 & 1 & \frac{\hat{g}_3 \alpha +\alpha +\hat{g}_2}{1-\hat{g}_3} \\
 0 & \frac{1}{\hat{g}_3+1} & \frac{-\hat{g}_3 \alpha +\alpha -\hat{g}_2}{\hat{g}_3+1} & \frac{\left(\hat{g}_2+\alpha  \hat{g}_3\right){}^2}{\hat{g}_3^2-1} \\
\end{array}
\right).  $$
Using the coordinate \tfn{}
\begin{align*}
s_1&=u,\\
\hat g_1&=\frac{\alpha^2}{2}u + v+\alpha\,z_3(\log(\cosh u)-1)- \frac{1}{2}
   \tanh u\left[\alpha^2\left((\log (\cosh u)-1)^2+{z_3}^2\right)\right]\\
\hat g_2&=z_3- \alpha \tanh u\log (\cosh u),\\
\hat g_3&=\tanh u
\end{align*}
for $|\hat g_3|<1$, the dual metric acquires the Brinkmann form
\begin{equation}\label{ppwS3,22a}
ds^2=-2\frac{z_3^2}{\cosh^2 u} du^2 + 2 du dv +dz_3^2 + dz_4^2.
\end{equation}
Changing $\cosh u\,\rightarrow\, \sinh u$ and $\tanh u\,\rightarrow\, \coth u$ in the transformation, we get
\begin{equation}\label{ppwS3,22b}
ds^2=2\frac{z_3^2}{\sinh^2 u} du^2 + 2 du dv +dz_3^2 + dz_4^2
\end{equation}
for $|\hat g_3|>1$. Torsion in both cases vanishes. This restores the result of \cite{php} for any initial torsionless $B$-field. The background appeared  already  in \cite{php,hp} by dualization with respect to subgroups generated by $S_{3,22}$, $S_{3,23}$, $S_{3,25}$, $S_{3,26}$, $S_{4,7}$, and $S_{4,8}$ without $B$-field.

\subsubsection*{Dualization with respect to subgroups generated by $S_{3,23},S_{3,26}$}

In these cases, a torsionless background on $M$ is obtained if the matrix $B(s_1)$ fulfills
$$B_2(s_1)=- B_4'(s_1){-}\beta\,B_3(s_1),\qquad B_3(s_1)=- B_5'(s_1),\qquad B_6(s_1)=0$$
where $\beta=0$ for $S_{3,26}$ and  $\beta\neq 0$ for $S_{3,23}$. We choose $\xi(s)=(\frac{\sigma}{2},0,\alpha \log |s_1|,\frac{\sigma}{2})$, ${\sigma=\text{sgn}(s_1)}$, ${s_1\neq 0}$. The sigma model is dualizable for $B_5(s_1)\neq \pm1$. The dual background $\wh\cf(s,\hat g)$ has rather complicated form with vanishing scalar curvature $\wh {\cal R}$ and torsion $\wh {\cal H}$. Dependence of the dual model on components of the matrix $B(s_1)$ can be eliminated by the \coor{} \tfn
\begin{equation*}
\hat{g}_1\mapsto\hat{g}_1+ \int^{s_1} B_1(r) \, dr, \qquad
\hat{g}_2\mapsto\hat{g}_2-B_4(s_1)+\beta\,B_5(s_1),\qquad
\hat{g}_3\mapsto\hat{g}_3-B_5(s_1)
\end{equation*}
that brings the dual background to the form
$$\wh\cf(s,\hat g)=\left(
\begin{array}{cccc}
 \frac{\alpha ^2}{s_1^2} & 0 & 0 & -\frac{\alpha ^2}{s_1 \left(\sigma-\hat{g}_3\right)} \\
 0 & 0 & 0 & \frac{1}{\sigma-\hat{g}_3} \\
 0 & 0 & 1 & \frac{\beta  \sigma+\hat{g}_2+\beta  \hat{g}_3}{\sigma-\hat{g}_3} \\
 \frac{\alpha ^2}{s_1 \left(\sigma+\hat{g}_3\right)} & \frac{1}{\sigma+\hat{g}_3} & -\frac{-\beta
   \sigma+\hat{g}_2+\beta  \hat{g}_3}{\sigma+\hat{g}_3} & -\frac{\alpha ^2+\left(\hat{g}_2+\beta \hat{g}_3\right)^2}{1-\hat{g}_3^2} \\
\end{array}
\right)$$
where $\alpha > 0$ for both subgroups. By rather complicated coordinate \tfn{} (see \cite{php}),
the dual metric acquires the Brinkmann form \eqref{ppwS3,22a} or \eqref{ppwS3,22b}. Torsion in both cases vanishes.

\subsection*{A.5\mez Dual with respect to subgroup generated by Bianchi $6_0$}\label{bia60}

\subsubsection*{Dualization with respect to subgroup generated by $S_{3,27}$}

In this case, a torsionless background on $M$ is obtained if the matrix $B(s_1)$ fulfills
$$B_2(s_1)=\csc\gamma\, B_5'(s_1), \qquad B_3(s_1)=\csc\gamma\, B_4'(s_1), \qquad B_6(s_1)=B_6=const.$$
where $0 < \gamma < \pi, \gamma \neq \frac{\pi}{2}$. We choose $\xi(s)=(0,0,\sqrt{s_1},0)$ with $B_4(s_1)^2-B_5(s_1)^2 + (B_6^2-1) s_1 \cos^2 \gamma \neq0$, $s_1 >0$. Dual scalar curvature $\wh {\cal R}$ vanishes for $B_6=\pm 1$ and we get pp-wave background with covariantly constant null vector field $\partial_{\hat g_1}$. Using the transformation of coordinates
\begin{align*}
s_1&=z_3^2+z_4^2,\\
\hat{g}_1&=-B_6 \,v \cos\gamma-\frac{\sin\gamma}{4} \, e^{2 u \tan\gamma}+\int^{z_3^2+z_4^2} B_1(r) \, dr,\\
\hat{g}_2&= e^{u \tan\gamma} \cosh \left(\tan(\gamma) \arctan\frac{z_4}{z_3}\right)+ \csc\gamma\,B_5\left(z_3^2+z_4^2\right),\\
\hat{g}_3&=-e^{u \tan \gamma} \sinh \left(\tan(\gamma) \arctan\frac{z_4}{z_3}\right)+\csc\gamma\,B_4\left(z_3^2+z_4^2\right),
\end{align*}
the dual background with $B_6=\pm 1$ acquires the Brinkmann form\footnote{For $B_6=-1$, it is necessary to interchange $\cosh $ and $\sinh$.} \eqref{hom-isotr-pp} with torsion \eqref{hom-isotr-H}.

\subsection*{A.6\mez Duals with respect to four-dimensional subgroups}\label{4dim}

For four-dimensional subgroups the matrix $B$ is constant. Since the conditions on the initial $B$-field
\begin{equation}
{\mathcal H}=d\mathcal B=0 \label{torsionless_4}
\end{equation}
differ case by case, we shall give the solutions when discussing particular dual backgrounds.

\subsubsection*{Dualization with respect to subgroups generated by $S_{4,7}$, $S_{4,8}$}

As we can see from Tab. \ref{table4}, non-isomorphic subalgebras $S_{4,7}$ and $S_{4,8}$ differ in the value of the parameter $\alpha$, which is positive for $S_{4,7}$ and vanishes for $S_{4,8}$. We choose $\xi(s) = (\frac{1}{2}, 0, 0,\frac{1}{2})$. In both cases the conditions \eqref{torsionless_4} hold when $B_4=B_5=B_6=0$.  The scalar curvatures $\wh{\mathcal{R}}$ as well as torsions $\wh{\mathcal{H}}$ then vanish. The symmetric parts of both backgrounds $\wh\cf$ can be brought to Brinkmann forms \eqref{ppwS3,22a} or \eqref{ppwS3,22b} that are independent of the parameter $\alpha$. As the formulas for $S_{4,7}$ are too complicated to display, we give only the results valid for $S_{4,8}$.

The symmetric part of dual background
$$\wh\cf(\hat g_1, \hat g_2, \hat g_3, \hat g_4)=\left(
\begin{array}{cccc}
 0 & 0 & -\frac{1}{B_2+\hat g_3-1} & 0 \\
 0 & 1 & -\frac{B_1+\hat g_2}{B_2+\hat g_3-1} & 0 \\
 \frac{1}{B_2+\hat g_3+1} & -\frac{B_1+\hat g_2}{B_2+\hat g_3+1} & \frac{B_1^2+2 \hat g_2 B_1+B_3^2+\hat g_2^2}{B_2^2+2 \hat g_3 B_2+{\hat g_3}^2-1} & -\frac{B_3}{B_2+\hat g_3+1} \\
 0 & 0 & -\frac{B_3}{B_2+\hat g_3-1} & 1 \\
\end{array}
\right)$$
can be brought to Brinkmann form \eqref{ppwS3,22a} using transformation
\begin{align*}
\hat{g}_1 &=v+\frac{1}{2} B_3^2 u-\frac{1}{2} z_3^2 \tanh (u), & \hat{g}_3 &=\tanh (u)-B_2,\\
\hat{g}_2 &=z_3-B_1, & \hat{g}_4 &=z_4-B_3 \log (\cosh (u))
\end{align*}
for $|\hat{g}_3+B_2|<1$, or to the form \eqref{ppwS3,22b} for $|\hat{g}_3+B_2|>1$ using similar transformation where we change $\tanh \rightarrow \coth$ and $\cosh \rightarrow \sinh$.

\subsubsection*{Dualization with respect to subgroup generated by $S_{4,17}$}

We choose $\xi(s) = (0, 0, 0, 0)$. When conditions \eqref{torsionless_4} hold, i.e. when $B_5=B_6=0$, the shift
$$\hat{g}_2\mapsto\hat{g}_2-B_2,\qquad  \hat{g}_3\mapsto\hat{g}_3+B_1$$
simplifies the dual background to the form
$$\wh\cf(\hat g_1, \hat g_2, \hat g_3, \hat g_4)=\left(
\begin{array}{cccc}
 0 & 0 & 0 & -\frac{1}{2 \epsilon +B_3} \\
 0 & \frac{1}{B_4^2+1} & -\frac{B_4}{B_4^2+1} & \frac{B_4 \hat{g}_2+\hat{g}_3}{(2 \epsilon +B_3) \left(B_4^2+1\right)} \\
 0 & \frac{B_4}{B_4^2+1} & \frac{1}{B_4^2+1} & \frac{B_4 \hat{g}_3-\hat{g}_2}{(2 \epsilon +B_3) \left(B_4^2+1\right)} \\
 \frac{1}{B_3-2 \epsilon } & \frac{B_4 \hat{g}_2-\hat{g}_3}{(2 \epsilon -B_3) \left(B_4^2+1\right)} & \frac{\hat{g}_2+B_4 \hat{g}_3}{(2 \epsilon -B_3) \left(B_4^2+1\right)} & -\frac{\hat{g}_2^2+\hat{g}_3^2}{\left(4 \epsilon ^2-B_3^2\right) \left(B_4^2+1\right)} \\
\end{array}
\right)$$
where $\epsilon=\pm 1$. Subsequent coordinate transformation
\begin{align*}
\hat{g}_1 &=v+\frac{B_4}{2} \left(z_3^2+z_4^2\right), & \hat{g}_3 &=\sqrt{B_4^2+1} \left(z_3 \sin \frac{B_3 u}{2 \epsilon }+z_4 \cos \frac{B_3 u}{2 \epsilon}\right),\\
\hat{g}_4 &=u \left(\frac{B_3^2}{2 \epsilon }-2 \epsilon \right), & \hat{g}_2 &=\sqrt{B_4^2+1} \left(z_3 \cos \frac{B_3 u}{2 \epsilon}-z_4 \sin \frac{B_3 u}{2 \epsilon }\right)
\end{align*}
brings the metric to Brinkmann form \eqref{hom-isotr-pp}. The corresponding torsion is \eqref{hom-isotr-H}.

\subsubsection*{Dualization with respect to subgroup generated by $S_{4,23}$}

We choose $\xi(s) = (0, 0, 0, 0)$. Conditions \eqref{torsionless_4} hold when $B_2=B_6=0$. After the shift
$$\hat{g}_3\mapsto\hat{g}_3-B_5,\qquad  \hat{g}_4\mapsto\hat{g}_4-B_1$$
the dual background is still too complicated to display, but its scalar curvature $\wh{\mathcal{R}}$ is given by
$$\wh{\mathcal{R}}=\frac{14 \left(B_4^4+B_4^2\right)}{\left(2 \alpha  B_4 \hat{g}_3+2 B_5 \hat{g}_3-2 B_4 \hat{g}_4+\hat{g}_3^2+\alpha ^2-B_3^2 B_4^2+B_5^2-2 B_1 B_4\right)^2},$$
where $\alpha\neq 0$. Scalar curvature vanishes for $B_4=0$, in which case the dual background reads
\begin{align*}
\wh\cf&(\hat g_1, \hat g_2, \hat g_3, \hat g_4)=\\
&\left(
\begin{array}{cccc}
 0 & 0 & 1 & 0 \\
 0 & \frac{1}{\alpha ^2+\hat{g}_3^2} & \frac{\alpha  B_5-\alpha  \hat{g}_3-B_3 \hat{g}_3+\hat{g}_4}{\alpha ^2+\hat{g}_3^2} & -\frac{\hat{g}_3}{\alpha ^2+\hat{g}_3^2} \\
 1 & -\frac{\alpha  B_5-\alpha  \hat{g}_3+B_3 \hat{g}_3+\hat{g}_4}{\alpha ^2+\hat{g}_3^2} & -\frac{\left(B_3^2+\left(B_5-\hat{g}_3\right)^2\right) \alpha ^2+2 \left(B_5-\hat{g}_3\right) \hat{g}_4 \alpha +\hat{g}_4^2}{\alpha ^2+\hat{g}_3^2} & \frac{\left(B_5-\hat{g}_3\right) \hat{g}_3 \alpha -\alpha ^2 B_3+\hat{g}_3 \hat{g}_4}{\alpha ^2+\hat{g}_3^2} \\
 0 & \frac{\hat{g}_3}{\alpha ^2+\hat{g}_3^2} & \frac{B_3 \alpha ^2+\left(B_5-\hat{g}_3\right) \hat{g}_3 \alpha +\hat{g}_3 \hat{g}_4}{\alpha ^2+\hat{g}_3^2} & \frac{\alpha ^2}{\alpha ^2+\hat{g}_3^2} \\
\end{array}
\right).
\end{align*}
Subsequent coordinate transformation
\begin{align*}
\hat{g}_1=&-\frac{v}{\alpha}+\frac{-12 \alpha ^2 B_3^2 u+\alpha ^4 u \left(3 u^4-28 u^2+48\right)}{24 \alpha }\\ &+2 \alpha \sqrt{u^2+1} z_2 -\frac{1}{2} \alpha u^2 \sqrt{u^2+1} z_2 +\frac{u \left(\left(u^2+2\right) z_2^2+z_3^2\right)}{2\alpha(u^2+1)},\\
\hat{g}_2=& \frac{1}{2} \alpha  \left(\alpha  B_3 u^2+2 \sqrt{u^2+1} z_3\right),\\
\hat{g}_3=&- \alpha  u,\\
\hat{g}_4=&-\alpha  B_5-\frac{1}{2} \alpha ^2 u \left(u^2-4\right)+\sqrt{u^2+1} z_2
\end{align*}
brings the metric to Brinkmann form \eqref{ppwS3,17}. The corresponding torsion is  \eqref{torS3,17}.

\subsubsection*{Dualization with respect to subgroup generated by $S_{4,25}$}

We choose $\xi(s) = (0, 0, 0, 0)$. Conditions \eqref{torsionless_4} hold when $B_5=B_6=0$. After the shift
$$\hat{g}_3\mapsto\hat{g}_3+\frac{B_1}{2},\qquad  \hat{g}_4\mapsto\hat{g}_4+B_2$$
the dual background is still too complicated to display, but its scalar curvature $\wh{\mathcal{R}}$ is given by
$$\wh{\mathcal{R}}=\frac{224 B_3^2}{\left(8 B_3 \hat{g}_3+4 \hat{g}_4^2-B_3^2 B_4^2+4 \epsilon ^2\right)^2},$$
where $\epsilon=\pm 1$. Scalar curvature vanishes for $B_3=0$, in which case the dual background reads
$$\wh\cf(\hat g_1, \hat g_2, \hat g_3, \hat g_4)=\left(
\begin{array}{cccc}
 \frac{1}{\epsilon ^2+\hat{g}_4^2} & 0 & \frac{\hat{g}_4}{\epsilon ^2+\hat{g}_4^2} & -\frac{2 \hat{g}_3+B_4 \hat{g}_4}{2 \left(\epsilon ^2+\hat{g}_4^2\right)} \\
 0 & 0 & 0 & -\frac{1}{2} \\
 -\frac{\hat{g}_4}{\epsilon ^2+\hat{g}_4^2} & 0 & \frac{\epsilon ^2}{\epsilon ^2+\hat{g}_4^2} & \frac{2 \hat{g}_3 \hat{g}_4-\epsilon ^2 B_4}{2 \left(\epsilon ^2+\hat{g}_4^2\right)} \\
 \frac{2 \hat{g}_3-B_4 \hat{g}_4}{2 \left(\epsilon ^2+\hat{g}_4^2\right)} & -\frac{1}{2} & \frac{B_4 \epsilon ^2+2 \hat{g}_3 \hat{g}_4}{2 \left(\epsilon ^2+\hat{g}_4^2\right)} & -\frac{\epsilon ^2 B_4^2+4 \hat{g}_3^2}{4 \left(\epsilon ^2+\hat{g}_4^2\right)} \\
\end{array}
\right).$$
Subsequent coordinate transformation
\begin{align*}
\hat{g}_1=&\frac{B_4 u^2 \epsilon^2}{4}+\epsilon  \sqrt{u^2+1} z_3, &
\hat{g}_3=&\sqrt{u^2+1} z_4,\\
\hat{g}_2=& -\frac{2v}{\epsilon}+\frac{u \left(4 \left(\left(u^2+2\right) z_4^2+z_3^2\right)-B_4^2 \left(u^2+1\right) \epsilon ^2\right)}{4 \left(u^2+1\right) \epsilon }, & \hat{g}_4=&u \epsilon
\end{align*}
brings the metric to Brinkmann form \eqref{ppwS3,17}. The corresponding torsion is  \eqref{torS3,17}.

\subsubsection*{Dualization with respect to subgroups generated by $S_{4,26}$, $S_{4,27}$}

As we can see from Tab. \ref{table4}, non-isomorphic subalgebras $S_{4,26}$ and $S_{4,27}$ differ only in the value of the parameter $\alpha$. Accordingly, the results of dualization and corresponding coordinate transformations can be written concisely in terms of $\alpha$, which is positive for $S_{4,26}$ and vanishes for $S_{4,27}$. We choose $\xi(s) = (\half, 0, 0, \half)$. For both subalgebras the conditions \eqref{torsionless_4} hold when $B_4=B_5=B_6=0$.  The scalar curvatures $\wh{\mathcal{R}}$ as well as torsions $\wh{\mathcal{H}}$ then vanish. The shift
$$\hat{g}_2\mapsto\hat{g}_2-B_1,\qquad \hat{g}_3\mapsto\hat{g}_3-B_2,\qquad  \hat{g}_4\mapsto\hat{g}_4-B_3$$
simplifies the dual background to the form
$$\wh\cf(\hat g_1, \hat g_2, \hat g_3, \hat g_4)=\left(
\begin{array}{cccc}
 0 & 0 & 0 & \frac{1}{1-\hat{g}_4} \\
 0 & 1 & 0 & \frac{-B_3 \alpha +\hat{g}_4 \alpha +\alpha +\hat{g}_2}{1-\hat{g}_4} \\
 0 & 0 & 1 & -\frac{\hat{g}_3}{\hat{g}_4-1} \\
 \frac{1}{\hat{g}_4+1} & \frac{B_3 \alpha -\hat{g}_4 \alpha +\alpha -\hat{g}_2}{\hat{g}_4+1} & -\frac{\hat{g}_3}{\hat{g}_4+1} & \frac{\hat{g}_2^2+2 \alpha  \left(\hat{g}_4-B_3\right) \hat{g}_2+\hat{g}_3^2+\alpha ^2 \left(B_3-\hat{g}_4\right)^2}{\hat{g}_4^2-1} \\
\end{array}
\right).$$
Subsequent coordinate transformation
\begin{align*}
\hat{g}_1 =& v+\frac{1}{2} \left(-\alpha  z_3 (2 B_3 u+\log (1-\tanh (u))+\log (\tanh (u)+1)+2)\right)\\ & -\frac{1}{8} \alpha ^2 \tanh (u) (2 B_3 u+\log (1-\tanh (u))+\log (\tanh (u)+1)+2)^2\\
& + \frac{1}{2} \alpha ^2 \left(B_3^2+1\right) u - \alpha ^2 B_3 \log (\cosh (u))-\frac{1}{2}\left(z_3^2+z_4^2\right) \tanh (u) ,\\
\hat{g}_2 =&z_3+\frac{1}{2} \alpha  \tanh (u) \left(2 B_3 u+\log (1-\tanh (u))+\log (\tanh (u)+1)\right),\\
\hat{g}_3 =& z_4,\\
\hat{g}_4 =& \tanh (u)
\end{align*}
valid for $|\hat{g}_4|<1$ brings the metric to $\alpha$-independent Brinkmann form
$$ds^2=-2\frac{z_3^2 + z_4^2}{\cosh^2 u} du^2 + 2 du dv +dz_3^2 + dz_4^2.$$
For $|\hat{g}_4|>1$ similar transformation with $\tanh \rightarrow \coth$ and $\cosh \rightarrow \sinh$ gives metric
$$ds^2=2\frac{z_3^2 + z_4^2}{\sinh^2 u} du^2 + 2 du dv +dz_3^2 + dz_4^2.$$
This restores the results of \cite{hp}.

\subsubsection*{Dualization with respect to subgroup generated by $S_{4,28}$}

We choose $\xi(s) = (\half, 0, 0, \half)$. When the conditions \eqref{torsionless_4} hold, i.e. when $B_4=B_5=B_6=0$,  the scalar curvature $\wh{\mathcal{R}}$ vanishes. We denote $\beta:=\tan (\gamma)$, where $0< \gamma < \pi, \gamma \neq \frac{\pi}{2}$, see Tab. \ref{table4}. The shift
$$\hat{g}_4\mapsto\hat{g}_4+\frac{B_3}{\beta}$$
simplifies the dual background to the form
\begin{align*}
\wh\cf&(\hat g_1, \hat g_2, \hat g_3, \hat g_4)=\\
& \left(
\begin{array}{cccc}
 0 & 0 & 0 & \frac{1}{\beta  \left(\hat{g}_4-1\right)} \\
 0 & 1 & 0 & \frac{B_1-\beta  \hat{g}_2+\hat{g}_3}{\beta  \left(\hat{g}_4-1\right)} \\
 0 & 0 & 1 & \frac{B_2-\hat{g}_2-\beta  \hat{g}_3}{\beta  \left(\hat{g}_4-1\right)} \\
 -\frac{1}{\hat{g}_4 \beta +\beta } & \frac{B_1-\beta  \hat{g}_2+\hat{g}_3}{\hat{g}_4 \beta +\beta } & \frac{B_2-\hat{g}_2-\beta  \hat{g}_3}{\hat{g}_4 \beta +\beta } & \frac{B_1^2+2 \left(\hat{g}_3-\beta  \hat{g}_2\right) B_1+B_2^2-2 B_2 \left(\hat{g}_2+\beta  \hat{g}_3\right)+\left(\beta ^2+1\right) \left(\hat{g}_2^2+\hat{g}_3^2\right)}{\beta ^2 \left(\hat{g}_4^2-1\right)} \\
\end{array}
\right).
\end{align*}
A very complicated subsequent transformation brings the metric to Brinkmann form
$$ds^2=-\left(\frac{1+2 \beta^2\, \text{sech}^2(u)}{\beta^2}\right)\left(z_3^2+z_4^2\right) du^2 + 2 du dv +dz_3^2 + dz_4^2$$
for $|\hat{g}_4|<1$, or
$$ds^2=-\left(\frac{1-2 \beta^2\, \text{csch}^2(u)}{\beta^2}\right)\left(z_3^2+z_4^2\right) du^2 + 2 du dv +dz_3^2 + dz_4^2$$
for $|\hat{g}_4|>1$. The corresponding torsion of this model is $$
\wh{\mathcal{H}}=\frac{2}{\beta} du\wedge dz_3 \wedge dz_4.$$
This restores the results of \cite{hp}.

\subsubsection*{Dualization with respect to subgroup generated by $S_{4,29}$}

We choose $\xi(s) = (-\half, 0, 0, -\half)$. When conditions \eqref{torsionless_4} hold, i.e. when $B_4=B_5=0$, the shift
$$\hat{g}_2\mapsto\hat{g}_2+B_1 \cot (\gamma),\qquad  \hat{g}_3\mapsto\hat{g}_3-B_3,\qquad  \hat{g}_4\mapsto\hat{g}_4+B_2$$
simplifies the dual background to the form
$$\wh\cf(\hat g_1, \hat g_2, \hat g_3, \hat g_4)=\left(
\begin{array}{cccc}
 0 & \frac{\cot (\gamma )}{\hat{g}_2+1} & 0 & 0 \\
 \frac{\cot (\gamma )}{1-\hat{g}_2} & \frac{\cot ^2(\gamma ) \left(\hat{g}_3^2+\hat{g}_4^2\right)}{\left(B_6^2+1\right) \left(\hat{g}_2^2-1\right)} & \frac{\cot (\gamma ) \left(B_6 \hat{g}_3-\hat{g}_4\right)}{\left(B_6^2+1\right) \left(\hat{g}_2-1\right)} & \frac{\cot (\gamma ) \left(\hat{g}_3+B_6 \hat{g}_4\right)}{\left(B_6^2+1\right) \left(\hat{g}_2-1\right)} \\
 0 & -\frac{\cot (\gamma ) \left(B_6 \hat{g}_3+\hat{g}_4\right)}{\left(B_6^2+1\right) \left(\hat{g}_2+1\right)} & \frac{1}{B_6^2+1} & -\frac{B_6}{B_6^2+1} \\
 0 & \frac{\cot (\gamma ) \left(\hat{g}_3-B_6 \hat{g}_4\right)}{\left(B_6^2+1\right) \left(\hat{g}_2+1\right)} & \frac{B_6}{B_6^2+1} & \frac{1}{B_6^2+1} \\
\end{array}
\right)$$
where $0< \gamma < \pi, \gamma \neq \frac{\pi}{2}$. For $|\hat{g}_2|<1$ subsequent coordinate transformation
\begin{align*}
\hat{g}_1 &=v+\frac{1}{2} \left(B_6 \left(z_3^2+z_4^2\right)\right), & \hat{g}_3 &=\sqrt{B_6^2+1} (z_3 \cos (f_\gamma(u))-z_4 \sin (f_\gamma(u)),\\
\hat{g}_2 &=\tanh (u \tan (\gamma )), & \hat{g}_4 &=\sqrt{B_6^2+1} (z_4 \cos (f_\gamma(u))+z_3 \sin (f_\gamma(u)))
\end{align*}
where
$$f_\gamma(u)=\cot (\gamma ) \log (\cosh (u \tan (\gamma )))$$
brings this pp-wave metric to Brinkmann form \eqref{hom-isotr-pp}. The corresponding torsion of this model is \eqref{hom-isotr-H}. For $|\hat{g}_2|>1$ similar transformation with $\tanh \rightarrow \coth$ and $\cosh \rightarrow \sinh$ again gives metric \eqref{hom-isotr-pp} and torsion \eqref{hom-isotr-H}.

\subsubsection*{Dualization with respect to subgroups generated by $S_{4,31}$, $S_{4,33}$}

As we can see from Tab. \ref{table4}, subalgebras $S_{4,31}$ and $S_{4,33}$ contain the parameter $\beta\neq 0$ and differ only in the value of the parameter $\alpha$. Accordingly, the results of dualization and corresponding coordinate transformations can be written in terms of $\alpha$, which is positive for $S_{4,33}$ and vanishes for $S_{4,31}$. We choose $\xi(s) = (-\half, 0, 0, -\half)$. For both subalgebras the conditions \eqref{torsionless_4} hold when $B_4=B_6=0$ and $B_5=B_2$. The scalar curvatures $\wh{\mathcal{R}}$ then vanish. After the shift
$$\hat{g}_3\mapsto\hat{g}_3-B_2, \qquad  \hat{g}_4\mapsto\hat{g}_4-B_3$$
the dual background gets the form
\begin{align*}
&\wh\cf(\hat g_1, \hat g_2, \hat g_3, \hat g_4)=\\
& \left(\begin{array}{cccc}
 0 & 0 & -\frac{1}{\hat{g}_3+1} & 0 \\
 0 & \frac{1}{\beta ^2+\hat{g}_3^2} & \frac{\alpha  \beta -B_1+\left(\hat{g}_3+1\right) \hat{g}_4}{\left(\hat{g}_3+1\right) \left(\beta ^2+\hat{g}_3^2\right)} & -\frac{\hat{g}_3+1}{\beta ^2+\hat{g}_3^2} \\
 \frac{1}{\hat{g}_3-1} & -\frac{\alpha  \beta +B_1+\left(\hat{g}_3-1\right) \hat{g}_4}{\left(\hat{g}_3-1\right) \left(\beta ^2+\hat{g}_3^2\right)} & \frac{B_1^2-2 \hat{g}_4 B_1+\alpha ^2 \hat{g}_3^2+\left(\beta ^2+1\right) \hat{g}_4^2-2 \alpha  \beta  \hat{g}_3 \hat{g}_4}{\left(\hat{g}_3^2-1\right) \left(\beta ^2+\hat{g}_3^2\right)} & \frac{\alpha  \beta  \left(\hat{g}_3+1\right)+B_1 \left(\hat{g}_3+1\right)-\left(\beta ^2+1\right) \hat{g}_4}{\left(\hat{g}_3-1\right) \left(\beta ^2+\hat{g}_3^2\right)} \\
 0 & \frac{\hat{g}_3-1}{\beta ^2+\hat{g}_3^2} & \frac{\alpha  \beta  \left(\hat{g}_3-1\right)-B_1 \left(\hat{g}_3-1\right)-\left(\beta ^2+1\right) \hat{g}_4}{\left(\hat{g}_3+1\right) \left(\beta ^2+\hat{g}_3^2\right)} & \frac{\beta ^2+1}{\beta ^2+\hat{g}_3^2} \\
\end{array}
\right).
\end{align*}
A very complicated subsequent transformation brings the metric to the following $\alpha$-independent Brinkmann form
$$ds^2=\left(\frac{4 \left(-2 \beta ^2+\left(\beta ^2+1\right) \cosh (2 u)-1\right)}{\left(\beta ^2+\left(\beta ^2+1\right) \cosh (2 u)-1\right)^2} z_3^2 -\frac{4 \beta ^2 \left(\beta ^2+\left(\beta ^2+1\right) \cosh (2 u)+2\right)}{\left(\beta ^2+\left(\beta ^2+1\right) \cosh (2 u)-1\right)^2} z_4^2\right) du^2$$ $$+ 2 du dv +dz_3^2 + dz_4^2$$
for $|\hat{g}_4|<1$, with corresponding torsion
$$\wh{\mathcal{H}}=\frac{4 \beta }{\beta ^2+\left(\beta ^2+1\right) \cosh (2 u)-1} du\wedge dz_3 \wedge dz_4,$$
or to the form
$$ds^2=\left(-\frac{4 \left(2 \beta ^2+\left(\beta ^2+1\right) \cosh (2 u)+1\right)}{\left(-\beta ^2+\left(\beta ^2+1\right) \cosh (2 u)+1\right)^2} z_3^2 +\frac{4 \beta ^2 \left(-\beta ^2+\left(\beta ^2+1\right) \cosh (2 u)-2\right)}{\left(-\beta ^2+\left(\beta ^2+1\right) \cosh (2 u)+1\right)^2} z_4^2\right) du^2$$ $$+ 2 du dv +dz_3^2 + dz_4^2$$
for $|\hat{g}_4|>1$, with corresponding torsion
$$\wh{\mathcal{H}}=\frac{4 \beta }{-\beta ^2+\left(\beta ^2+1\right) \cosh (2 u)+1} du\wedge dz_3 \wedge dz_4.$$
Although it is not obvious at first sight, this restores the results of \cite{hp}.



\begin{thebibliography}{99}
\bibitem{buscher:ssbfe} T. H. Buscher, \emph{A symmetry of the String Background Field Equations}, {Phys. Lett. B} 194 (1987) 51.

\bibitem{delaossa1992}
X. C. de~la Ossa and F.~Quevedo, {\it Duality symmetries from non-abelian isometries in string theories},  Nucl. Phys. {B 403} (1993) 377, [hep-th/9210021].
  
\bibitem{klise}{C. Klim\v c\'ik and P. \v Severa, {\emph{Dual non-Abelian duality and the Drinfeld double}}, Phys. Lett. B 351 (1995) 455, [hep-th/9502122].}

\bibitem{CavGual}
G. Cavalcanti and M. Gualtieri, \emph{Generalized complex geometry and T-duality}, In: A Celebration of the Mathematical Legacy of Raoul Bott (CRM Proceedings \& Lecture Notes), American Mathematical Society  (2010) 341, [arXiv:1106.1747].

\bibitem{severa2015}
{P. {\v{S}}evera, {\it Poisson--Lie T-Duality and Courant Algebroids}, Lett. in Mat. Phys. 105 (12) (2015) 1689, [arXiv:1502.04517].}

\bibitem{Bouwknegt2017}
{P. Bouwknegt, M. Bugden, C. Klim{\v{c}}{\'i}k, and K. Wright, {\it Hidden isometry of ``T-duality without isometry''}, JHEP 2017 (8) (2017) 116, [arXiv:1705.09254].}

\bibitem{Hassan} S. F. Hassan, \emph{T-Duality, Space-time Spinors and R-R Fields in Curved Backgrounds}, Nucl. Phys. B 568 (2000) 145, [hep-th/9907152].

\bibitem{1012.1320 sfethomp}  K.~Sfetsos and D. C.~Thompson, {\it On non-abelian T-dual geometries with Ramond fluxes}, Nucl. Phys.  {B846 } (2011)  21, [arXiv:1012.1320].

\bibitem{1104.5196 LCST}  Y.~Lozano, E.~O'Colgain, K.~Sfetsos, and D.~C.~Thompson,   {\it Non-abelian T-duality, Ramond fields and coset geometries}, JHEP {1106} (2011) 106, [arXiv:1104.5196].

\bibitem{Itsios:2013wd} G.~Itsios, C.~Nunez, K.~Sfetsos, and D.~C.~Thompson, \emph{Non-Abelian T-duality and the AdS/CFT correspondence:new N=1 backgrounds}, Nucl.\ Phys.\ B {873}, 1 (2013), {[arXiv:1301.6755]}.

\bibitem{SadSheikh} D. Sadri, M.M. Sheikh-Jabbari, \emph{The Plane-Wave/Super Yang-Mills Duality}, Rev.Mod.Phys. 76 (2004) 853, [hep-th/0310119].
 

\bibitem{BorWulff1} R. Borsato, L. Wulff, \emph{Integrable deformations of T-dual {$\sigma$}-models}, Phys. Rev. Lett. 117, 251602 (2016), [arXiv:1609.09834].

\bibitem{BorWulff} R. Borsato, L. Wulff, \emph{On non-abelian T-duality and deformations of supercoset string sigma-models}, JHEP 2017 (24) (2017), [arXiv:1706.10169].

\bibitem{php} F. Petr\'asek, L. Hlavat\'y, and I. Petr, \emph{Plane-parallel waves as duals of the flat background II: T-duality with spectators}, Class. Quantum Grav. 34 (2017) 155003, [arXiv:1612.08015].

\bibitem{hp} L. Hlavat\'y and I. Petr, \emph{Plane-parallel waves as duals of the flat background}, Class. Quantum Grav. 32 (2015) 035005, [arXiv:1406.0971].

\bibitem{PWZ} J. Patera, R. T. Sharp, P. Winternitz, and H. Zassenhauss,
{\emph {Subgroups of the Poincar\'e group and their invariants}}, J.
Math. Phys. 17 (1976) 977.

\bibitem{AAG:global} E. Alvarez, L. Alvarez-Gaume, J. L. F. Barbon, and Y. Lozano, \emph{ Some Global Aspects of Duality is String Theory}, Nucl.Phys. B415 (1994) 71, [hep-th/9309039].


\bibitem{tsey:revExSol}{A. A. Tseytlin, {\emph {Exact solutions of closed string theory}},
Class. Quant. Grav. 12 (1995) 2365, [hep-th/9505052].}

\bibitem{tseyt94} A. A. Tseytlin, {\it Exact string solutions and duality}, Proc. Cosmology, Paris (1994) 371, [hep-th/9407099].

%

\bibitem{Sfetseyt94} K. Sfetsos and A. A. Tseytlin, \emph{Four Dimensional Plane Wave String Solutions with Coset CFT Description}, Nucl. Phys.
B 427 (1994) 245, [hep-th/9404063].

\bibitem{BLPT} M. Blau, M. O'Loughlin, G. Papadopoulos, and A.  A. Tseytlin, \emph{Solvable models of strings in homogeneous plane wave backgrounds}, Nucl. Phys. B 673 (2003) 57, [hep-th/0304198].

\bibitem{papa} G. Papadopoulos, J. G. Russo, and A. A. Tseytlin, \emph{Solvable model of strings in a time-dependent plane-wave background}, Class. Quantum
Grav. 20 (2003) 969, [hep-th/0211289].

\bibitem{1205.2274 ILCS} G. Itsios, Y. Lozano, E. O'Colgain, and K. Sfetsos, {\it Non-Abelian T-duality and consistent truncations in type-II supergravity}, JHEP 1208 (2012) 132, [arXiv:1205.2274].

\bibitem{DMRV}{H. Dimov}, {S. Mladenov}, {R. C. Rashkov}, and {T. Vetsov} \emph{Non-abelian T-duality of Pilch-Warner background}, Fortschr. Phys. 64, 657 (2016), [arXiv:1511.00269].

\bibitem{LNZ} Y. Lozano, C. Nunez, and S. Zacarias, \emph{BMN Vacua, Superstars and Non-Abelian T-duality}, JHEP 2017 (0) (2017), [arXiv:1703.00417].

\bibitem{hp:uniq} L. Hlavat\'y and F. Petr\'asek, \emph{On
uniqueness of T-duality with spectators}, Int. J. Mod. Phys. A 31
(2016) 1650143, [arXiv:1606.02522].

\bibitem{blauffpapa}{M. Blau, J. Figueroa-O'Farrill, and G. Papadopoulos,
 \emph{Penrose limits, supergravity and brane dynamics
}, Class. Quantum Grav. 19 (2002) 4753, [hep-th/0202111].}

\bibitem{Penrose} R. Penrose, {\it Any Space-Time has a Plane Wave as a Limit},
in: Differential Geometry and Relativity, vol. 3, eds. M. Cahen and
M. Flato, Reidel, Dordrecht 1976.

\bibitem{Gueven} R. Gueven, {\it Plane wave limits and T-duality},
Phys. Lett. B 482 (2000), 255 - 263, [hep-th/0005061v1].

\bibitem{AntonObers} I. Antoniadis and N.A. Obers, \emph{Plane Gravitational Waves in String Theory}, Nucl. Phys. B423 (1994) 639, [hep-th/9403191].

\end{thebibliography}
\end{document}